\documentclass[fp,twocolumn]{jpsj3}
\usepackage{txfonts}

\usepackage{color}
\usepackage{ulem}

\newcommand{\Nfave}{N_{f, \text{ave}}}
\newcommand{\Nfmin}{N_{f, \text{min}}}

\title{Qubit-efficient embedding of parity-encoded Hamiltonians in quantum annealers}

\author{Ryoji Miyazaki$^{1, 2}$}
\inst{$^1$Secure System Platform Research Laboratories, NEC Corporation, Kawasaki 211-8666, Japan \\
$^2$NEC-AIST Quantum Technology Cooperative Research Laboratory, National Institute of Advanced Industrial Science and Technology
(AIST), Tsukuba, Ibaraki 305-8568, Japan} 

\abst{
The Sourlas-Lechner-Hauke-Zoller (SLHZ) scheme for quantum annealing uses the parity to encode logical variables 
and has several advantages, 
but it has not been implemented for large-scale quantum annealers.
If the SLHZ-based approach can be implemented on currently available quantum annealers, 
we can evaluate its performance.
An efficient method to embed the parity-encoded model into the hardware graphs of available quantum annealers is one of the key elements for this approach.
We propose a qubit-efficient embedding scheme for parity-encoded Hamiltonians on quantum annealers with the Zephyr connectivity.  
We give an explicit constructive embedding of the interaction graph of an intermediate Hamiltonian, 
which contains only one- and two-body interactions, 
into the Zephyr graph.
Our embedding maps each spin to a two-qubit chain using systematic chain-assignment rules. 
Its validity is verified via the resulting chain-to-chain connectivity. 
Our embedding also provides practical flexibility. 
Chains assigned to ancillary spins allow reduction to a single physical qubit, 
leading to options to avoid inactive qubits. 
The number of required qubits per spin in the parity Hamiltonian is three, 
which is fewer than that for a known embedding scheme for the Pegasus graph.  
}

\newcommand{\alphabar}{\bar{\alpha}}
\newcommand{\betabar}{\bar{\beta}}

\begin{document}
\maketitle

\section{Introduction}

Quantum annealing~\cite{T.Kadowaki1998, E.Farhi2000, T.Albash2018Jan, P.Hauke2020} is a heuristic approach 
that utilizes quantum effects to solve combinatorial optimization problems.
In the standard formulation, a combinatorial optimization problem is mapped to an Ising model
whose ground state represents the solution of the original problem~\cite{A.Lucas2014}.
Quantum annealing attempts to reach this ground state by gradually decreasing quantum fluctuations such as transverse
fields. 

A widely accessible platform for quantum annealing 
includes devices with 4400 qubits on the Zephyr topology~\cite{K.Boothby2021} 
and 5612 qubits on the Pegasus topology~\cite{K.Boothby2020arXiv}.
One practical limitation is that the connectivity between physical qubits is restricted by the device topology.
Consequently, 
couplings in an Ising Hamiltonian representing a problem to be solved 
cannot always be mapped directly 
to couplings available on the hardware graph, 
and one typically requires the so-called minor embedding~\cite{V.Choi2008, V.Choi2011}.
In minor embedding, 
a logical variable is represented by a chain of physical qubits 
that behave as a single effective qubit in the lowest-energy manifold due to strong intra-chain coupling.
One then constructs a graph consisting of these chains 
so that the couplings of the logical problem can be realized by couplers between chains.
While applications of minor embedding are known for some highly symmetric problems 
such as clique problems~\cite{K.Boothby2020arXiv, K.Boothby2021}, 
and heuristic approaches also exist~\cite{J.Cai2014arXiv}, 
embedding overhead remains a bottleneck for scaling up problem sizes on available hardware.

Concerning the embedding, 
the Sourlas-Lechner-Hauke-Zoller (SLHZ) scheme~\cite{N.Sourlas2005, W.Lechner2015} is different from the conventional one~\cite{M.Johnson2011, V.Choi2008, V.Choi2011}.
Instead of embedding the logical Ising Hamiltonian directly into the hardware graph via minor embedding, 
the SLHZ approach maps the logical Hamiltonian to 
an Ising Hamiltonian with multi-body interactions, 
where spins correspond to couplings of logical variables.
We call this Hamiltonian parity Hamiltonian and give its definition later.
Quantum annealing is then performed using the parity Hamiltonian, 
or using a corresponding Hamiltonian 
embedded in a layout of physical qubits with appropriate quantum driving terms.
A key advantage in the SLHZ framework is that 
it does not require couplings between distant qubits.
This locality is attractive for physical implementations 
because it leads to effective all-to-all logical connectivity using only local interactions.
A generalized version of the SLHZ scheme~\cite{K.Ender2023, M.Drieb-schon2023, M.Fellner2023, R.Hoeven2024} is also known.
By utilizing both four- and three-body interactions according to the problem instance, 
this generalized approach can embed problems more efficiently.

Despite these theoretical advantages, 
the SLHZ scheme has not been implemented in currently available quantum annealers.
The direct reason is that it requires four- or three-body couplings of physical qubits.
Theoretical ideas to implement the scheme have been proposed for superconducting circuits~\cite{M.Leib2016, N.Chancellor2017, S.Puri2017Jun, P.Zhao2018, R.Miyazaki2025} and Rydberg atoms~\cite{A.Glaetzle2017, C.Dlaska2022}, 
and a recent experiment has implemented a building block of the scheme, i.e., four-body interaction of qubits~\cite{Y.Kawakami2025arXiv}.
However, without a way to run large instances on existing large-scale annealers, 
it remains difficult to evaluate the performance of the scheme for large-scale problems.
If we have an efficient method to embed the parity Hamiltonian in the graphs of currently available quantum annealers, 
we can perform quantum annealing 
based on the SLHZ scheme using those annealers 
and evaluate its performance, in particular for solving relatively large problems.

One study~\cite{M.Cattelan2025} proposed a method to embed the parity Hamiltonian in the Pegasus graph.
In that approach, 
the model is mapped to a quadratic model with ancillary spins.
The embedding uses four or two physical qubits for each original spin and two physical qubits for each ancillary spin.
The study proposed two embedding variants that differ in the number of qubits used, 
but did not show a definite difference in the resulting quantum annealing performance, 
suggesting that the variant using fewer qubits can obtain results more qubit-efficiently at least.
This point is crucial because available devices allow only limited problem scales even with thousands of physical qubits.

To further extend this approach, 
we propose a method to efficiently embed the parity Hamiltonian in the Zephyr graph.
The method leverages the symmetry of the device graph and the parity Hamiltonian.
In terms of efficiency, the number of qubits required by the proposed method is three per spin in the parity Hamiltonian.

The next section introduces the Hamiltonians and graphs investigated in this paper.
We define the parity Hamiltonian for the SLHZ scheme, 
which contains four-body interactions, 
and then transform it into an intermediate Hamiltonian 
that has only one- and two-body terms.
We also describe the Zephyr graph.
Section III describes how we embed the intermediate Hamiltonian in the Zephyr graph. 
We introduce chains used in the embedding, 
present the rule for mapping spins in the intermediate Hamiltonian to the chains in the Zephyr graph, 
and discuss the optionality of the embedding.
The present study is summarized in Sec.\ IV.


\section{Hamiltonians and graphs}

\subsection{Parity Hamiltonian}

\begin{figure}[t]
\includegraphics[width=\columnwidth]{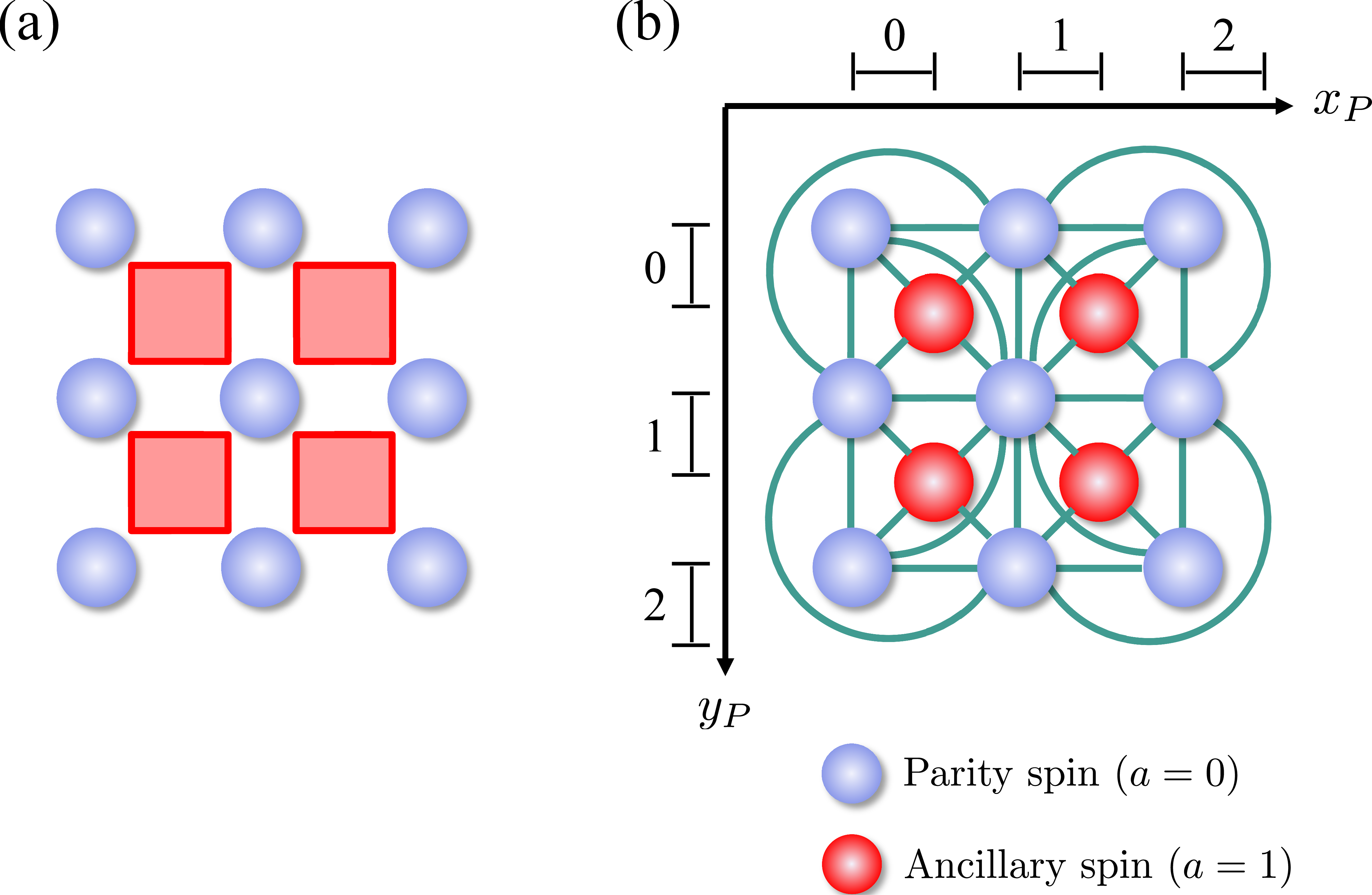}%
\caption{
(a) Lattice for the Parity Hamiltonian for $3 \times 3$ spins depicted as circles.
Squares represent four-body interactions between spins located at their vertices
and also plaquettes.
(b) Lattice for the corresponding intermediate Hamiltonian.
Each vertex on the lattice is parameterized by a coordinate $(x_P, y_P, a)$, 
where $a=0$ for parity spins and $a=1$ for ancillary spins.
A parity spin interacts with up to eight parity spins and four ancillary spins 
via two-body interactions depicted as edges.
Ancillary spins interact only with parity spins.}
\label{fig:LHZ-intermediate_lattice}
\end{figure}

We assume that 
the logical Hamiltonian representing the optimization problem is an Ising Hamiltonian~\cite{A.Lucas2014}.
Such logical Hamiltonians can be encoded into a parity Hamiltonian shown below.
We do not discuss the detailed procedure of the encoding~\cite{W.Lechner2015, K.Ender2023, R.Hoeven2024} here. 
We consider only the classical part of the parity Hamiltonian, written as
\begin{equation}
H = - \sum_i J_i \sigma_i - \sum_p C_p \prod_{i\in I_p} \sigma_i.
\label{eq:H_parity}
\end{equation}
Here, $\sigma_i$ represents an Ising spin on vertex $i$.
For simplicity of notation, 
we use it to represent $\sigma_i^{z}$ in the classical Hamiltonian.
The coefficients $J_i$ are usually coupling constants derived from the logical Hamiltonian. 
$I_p$ is the set of indices $i$ associated with plaquette $p$.
A plaquette consists of the spins in the minimal unit of lattice.
The four spins located at the vertices of a square in Fig.~\ref{fig:LHZ-intermediate_lattice}(a) form an example of plaquette.
Four-body interaction terms through the product over the spins belonging to the plaquette appear with positive coefficients $C_p$ 
and are leveraged to decode low-energy states of the parity Hamiltonian into states representing the logical Hamiltonian.
We also do not discuss the details of the decoding procedure~\cite{W.Lechner2015, K.Tiurev2025arXiv, Y.Nambu2026}.

We assume that the global shape of the lattice for the parity Hamiltonian is square 
as shown in Fig.~\ref{fig:LHZ-intermediate_lattice}(a).
Other shapes, such as triangles, can be treated as subgraphs of the square lattice and are
therefore included within the same framework.
The triangle case is considered in Sec.~\ref{sec:optionality}.

\subsection{Intermediate Hamiltonian}
\label{sec:intermediate_Hamiltonian}

Four-body interaction terms contained in the parity Hamiltonian $H$ [Eq.~(\ref{eq:H_parity})] cannot be directly mapped in 
graphs of currently available quantum annealers, 
since the graphs only have two-qubit couplings.
Therefore, we map the parity Hamiltonian to a Hamiltonian consisting only of one- and two-body terms
based on previous studies~\cite{A.Rocchetto2016, M.Cattelan2025}.
This subsection reviews the known procedure of the mapping~\cite{A.Rocchetto2016, M.Cattelan2025} for convenience.

We consider a single plaquette Hamiltonian for the four-spin case,
\begin{equation}
H^{(4)}_p = - \sum_{i=1}^{4} J_i \sigma_i - C_p \prod_{i=1}^{4} \sigma_i ,
\end{equation}
which contains the four-body product $\prod_{i=1}^{4}\sigma_i$.
We first perform the gauge transformation $\sigma_i \to \tilde{\sigma}_i$ and $J_i \to \tilde{J}_i$ that changes the sign of a spin and the corresponding $J_i$.
Assuming that $\sigma_1$ is the spin whose sign is changed, 
we have
\begin{equation}
\tilde{\sigma}_1,\tilde{\sigma}_2,\tilde{\sigma}_3,\tilde{\sigma}_4
= -\sigma_1,\sigma_2,\sigma_3,\sigma_4 ,
\end{equation}
\begin{equation}
\tilde{J}_1,\tilde{J}_2,\tilde{J}_3,\tilde{J}_4
= -J_1,J_2,J_3,J_4 .
\end{equation}
Hence the Hamiltonian becomes
\begin{equation}
H^{(4)}_p \to \tilde{H}^{(4)}_p
= - \sum_{i=1}^{4} \tilde{J}_i \tilde{\sigma}_i
+ C_p \prod_{i=1}^{4} \tilde{\sigma}_i .
\end{equation}
In lattices with multiple plaquettes, 
only one spin in each plaquette changes its sign in the gauge transformation.
We need to carefully determine which spin changes its sign, since each spin belongs to multiple plaquettes in the lattice.

We next introduce 
an ancillary spin $\tilde{\sigma}_0$ to construct
an effective Hamiltonian containing only one- and two-body terms that corresponds to $\tilde{H}^{(4)}_p$, 
\begin{equation}
H^{\prime(4)}_p
= - \sum_{i=1}^{4} \tilde{J}_i \tilde{\sigma}_i
+ \frac{C_p}{2} \left( \sum_{i=1}^{4} \tilde{\sigma}_i \right)^2
+ 2 C_p \tilde{\sigma}_0 \sum_{i=1}^{4} \tilde{\sigma}_i
+ C_p.
\end{equation}
Hereafter, 
we call $\tilde{\sigma}_i$ for $i=1,2,3,4$ parity spins
to distinguish them with the ancillary spin.
When the value of the ancillary spin is chosen to decrease the energy, $H^{\prime(4)}_p$ takes the same energy as that of $\tilde{H}^{(4)}_p$ for the same gauge-transformed parity-spin configuration.
By applying this construction to all plaquettes, the effective Hamiltonian for the whole system is
\begin{equation}
H'
= - \sum_i \tilde{J}_i \tilde{\sigma}_i
+ \sum_p C_p
\left[
\frac{1}{2}
\left(
\sum_{i \in I_p} \tilde{\sigma}_i
\right)^2
+ 2 \sigma_0 \sum_{i \in I_p} \tilde{\sigma}_i
\right],
\label{eq:H_intermediate}
\end{equation}
where the constant has been dropped.
We call $H'$ the intermediate Hamiltonian and will embed in the Zephyr graph.

Let $G$ be a graph, where each spin in $H'$ sits on a vertex of $G$, and each coupling of spins in $H'$ is represented by an edge of $G$.
We show $G$ in Fig.~\ref{fig:LHZ-intermediate_lattice}(b).
Each vertex in $G$ is represented by coordinate $(x_P,y_P,a)$: 
$a=0$ for parity spins and $a=1$ for ancillary spins.
Ignoring the boundaries, the number of the ancillary spins is equal to that of the parity spins.

We mainly consider the intermediate Hamiltonian $H'$ for the four-body-coupling plaquettes to be embedded.
Embedding the three-body case can be achieved by turning off the coupling to the physical qubits corresponding to the reduced one from the four-body case and adjusting the coupling constant.


\subsection{Zephyr graph}

\begin{figure}
\includegraphics[width=\columnwidth]{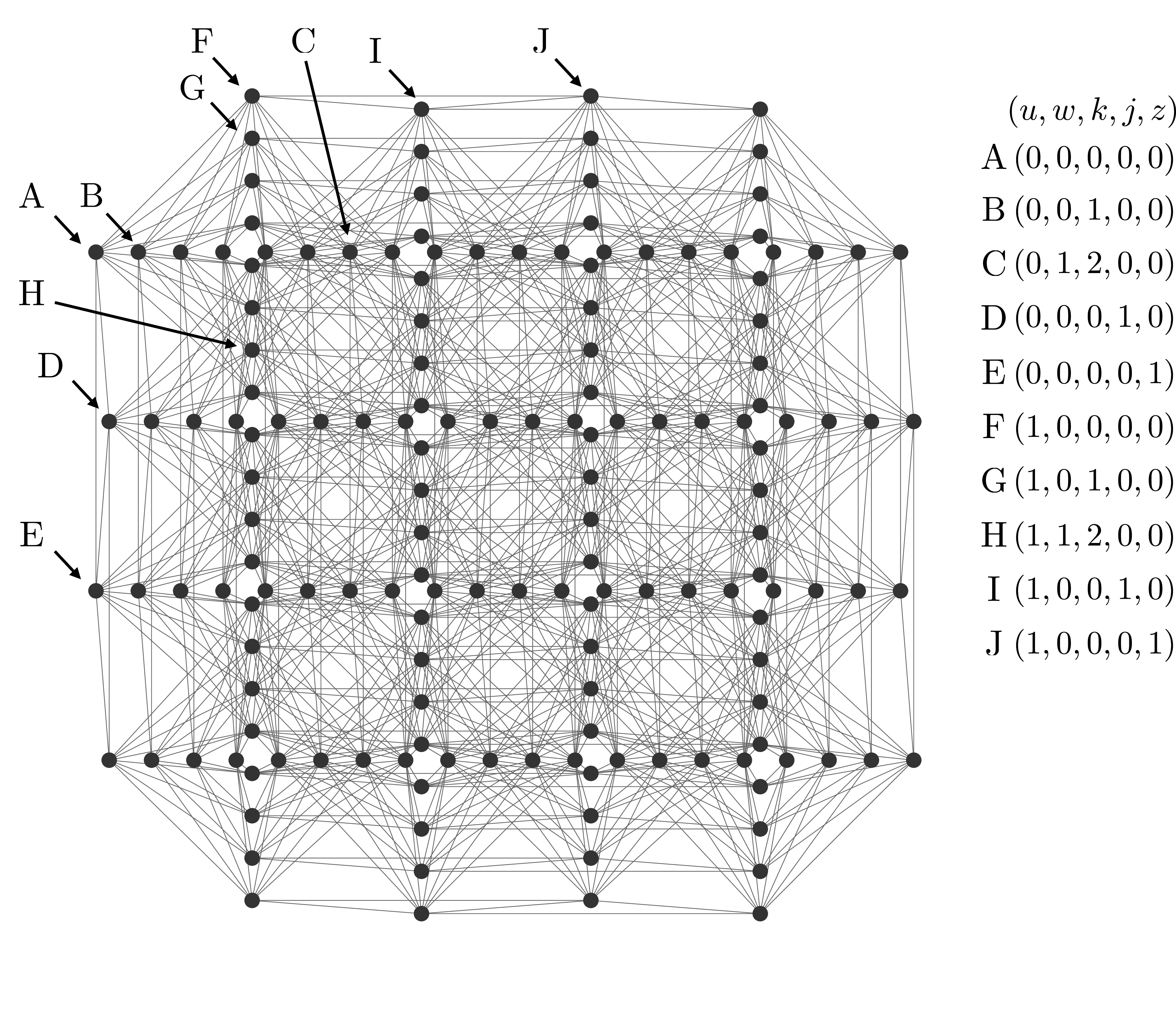}%
\caption{\label{fig:Zephyr}
			Zephyr graph $Z_{m, t}$ for $m=2$ and $t=4$
			where $m$ represents the number of units in each orientation, 
			and $t$ the number of qubits in a tile.
			We only consider $t=4$ and simply write it as $Z_2$. 
			Each vertex (qubit) is parametrized by $(u, w, k, j, z)$.
			$u$ indicates the orientation by 0 or 1, 
			$w$ ($z$) the index of the qubit's tile in the orientation perpendicular (parallel) to $u$, 
			$k$ the index of a qubit within a tile.
			They are in $0 \le w < 2m+1$, $0 \le z < m$, and $0 \le k < t$.
			Each qubit is shifted to the left/top for $j = 0$ and the right/bottom for $j = 1$.
			Parameters of qubits A--J are shown as examples.
			See the reference~\cite{K.Boothby2021} for the detailed description.}
\end{figure}

The Zephyr graph~\cite{K.Boothby2021} represents the physical layout of qubits and the availability of couplers in a quantum annealing processor.
Physical qubits sit on vertices of the graph, and the implemented couplings of qubits are represented by the edges.
Importantly, there are only two-qubit couplings. 
This restriction is precisely why, in Sec.~\ref{sec:intermediate_Hamiltonian}, 
we have introduced an intermediate Hamiltonian that contains only one- and two-body terms.
The Zephyr graph with grid parameter $m$ and tile parameter $t$ is denoted by $Z_{m, t}$, 
where $m$ represents the number of units in each orientation, 
and $t$ the number of qubits in a tile.
These two parameters characterize the size of the graph, 
and $Z_{m, t}$ contains $8tm^2 + 4tm$ qubits.
Each vertex is parametrized by a coordinate $(u,w,k,j,z)$~\cite{K.Boothby2021}.
$Z_{2,4}$ and its coordinate description are shown in Fig.~\ref{fig:Zephyr}.
See the reference~\cite{K.Boothby2021} for the detailed description.
We restrict our attention to the Zephyr family $Z_{m,t}$ for $t=4$~\cite{K.Boothby2021}, 
which is actually implemented in available devices, 
and simply represent it as $Z_m$.
Although we focus on $t=4$, 
we note that the embedding scheme described below works for $t \ge 3$.


\section{Embedding the intermediate Hamiltonian in the Zephyr graph}



Finding an embedding of the intermediate Hamiltonian $H'$ [Eq.~(\ref{eq:H_intermediate})] on the Zephyr graph $Z_m$ [Fig.~\ref{fig:Zephyr}]
is equivalent to finding a mapping
\begin{equation}
\phi : G \rightarrow Z_m ,
\end{equation}
such that each vertex of $G$ [Fig.~\ref{fig:LHZ-intermediate_lattice}(b)] is mapped 
to either a single vertex of $Z_m$ or a connected set of vertices of $Z_m$, 
which we call a chain.
Furthermore, for every edge between vertices $i$ and $j$ of $G$, 
corresponding to a required coupling between spins $i$ and $j$ in $H'$, 
there must exist at least one physical coupler between the chains assigned to $i$ and $j$ in the Zephyr graph.
This condition guarantees that all logical couplings in the intermediate Hamiltonian
can be implemented using the physical couplers available on the device.
Under such a mapping $\phi$, spins in $H'$ are implemented by chains in $Z_m$.
We focus on the construction of such a mapping and on the assignment
of chains to spins.
We do not consider the tuning of coupling constants in the physical Hamiltonian on
$Z_m$, i.e., the parameter setting problem~\cite{V.Choi2008}.



\subsection{Chains}

In our embedding scheme, each spin in the intermediate Hamiltonian $H'$ is mapped
to a two-qubit chain in the Zephyr graph $Z_m$.
We represent a chain consisting of two physical qubits, labeled $0$ and $1$, as
\[
[(u_0,w_0,k_0,j_0,z_0)\sim (u_1,w_1,k_1,j_1,z_1)].
\]
The two-qubit chains used in our embedding are specified as
\begin{align*}
&[(0,2z_1+j_1+\alpha,2-2\alpha+\beta,j_0,z_0)
\\
&\sim
(1,2z_0+j_0+\alpha,2-2\alpha+\beta,j_1,z_1)],
\end{align*}
where $\alpha,\beta = 0$ or $1$.
We later specify which chain is assigned to each spin in $H'$.
The above expression follows the conventional coordinates of the Zephyr graph,
but it is somewhat verbose.

\begin{figure}
\includegraphics[width=\columnwidth]{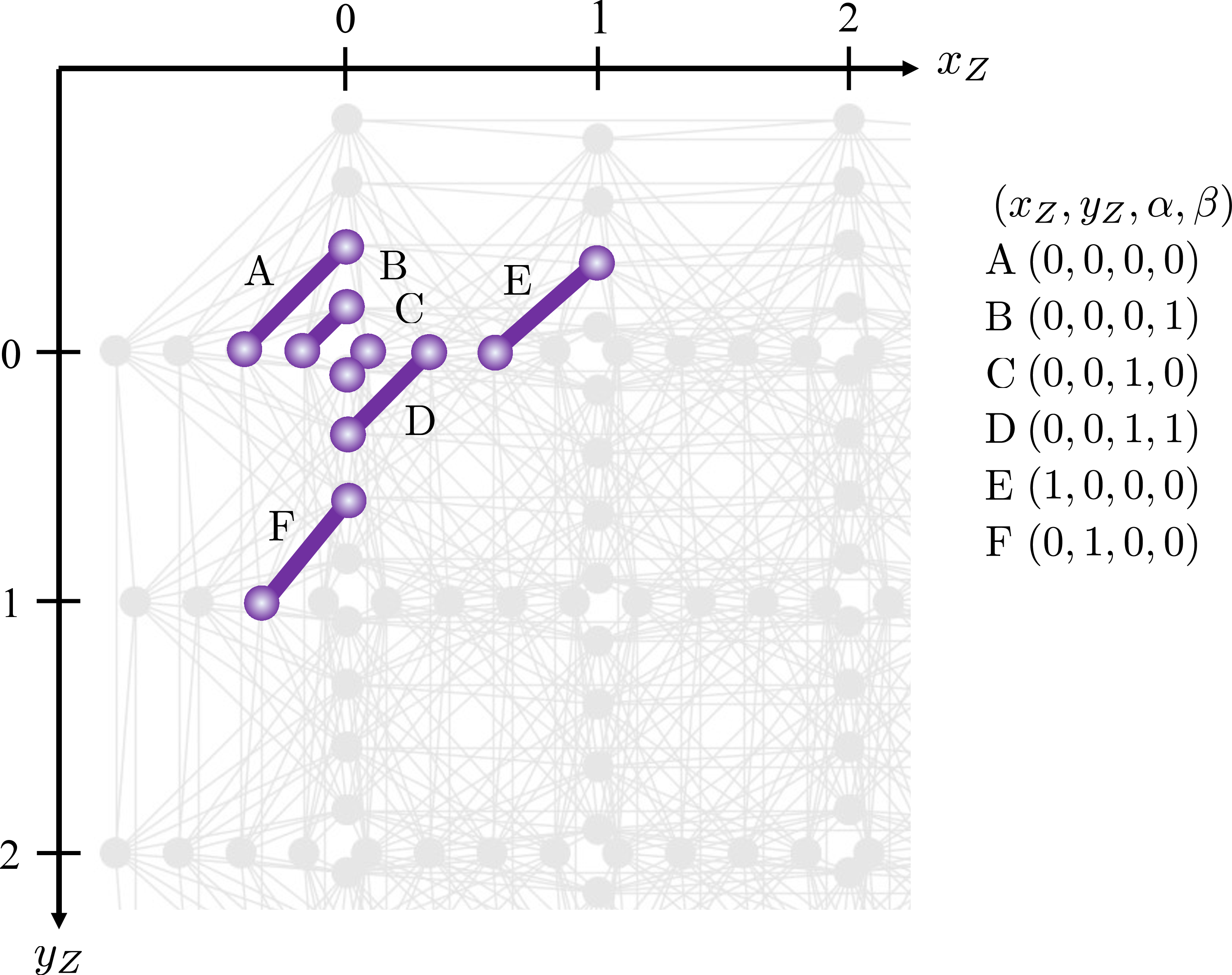}%
\caption{\label{fig:chains}
			Examples of chains specified by Eq.~(\ref{eq:chain}).
			Each chain consists of two vertices for $u = 0$ and 1 [Fig.~\ref{fig:Zephyr}] 
			and the edge connecting them.
			Vertices and edges for the labelled chains are highlighted.
			}
\end{figure}

For practical use, the same chain can also be written in terms of auxiliary integer
coordinates $x_Z$ and $y_Z$ as
\begin{equation}
\begin{split}
&\Bigg[ 
\left( 
0, x_Z + \alpha, 2 - 2 \alpha + \beta, y_Z - 2 \left\lfloor \frac{y_Z}{2} \right\rfloor, \left\lfloor \frac{y_Z}{2} \right\rfloor 
\right)
\\
& \sim
\left( 
1, y_Z + \alpha, 2 - 2 \alpha + \beta, x_Z - 2 \left\lfloor \frac{x_Z}{2} \right\rfloor, \left\lfloor \frac{x_Z}{2} \right\rfloor 
\right) 
\Bigg], 
\end{split}
\end{equation}
where 
\begin{equation}
\left\lfloor \frac{x_Z}{2} \right\rfloor = n,
\quad \text{s.t. } n \le \frac{x_Z}{2} < n+1 ,
\end{equation}
for integer $n$.
The auxiliary coordinates $x_Z$ and $y_Z$ are related to the original Zephyr parameters
by
\begin{equation}
x_Z = 2z_1 + j_1 ,
\end{equation}
and
\begin{equation}
y_Z = 2z_0 + j_0 .
\end{equation}
They satisfy $0 \le x_Z, y_Z \le 2m - 1$,
where $m$ characterizes the size of the Zephyr graph $Z_m$.
Note that
$x_Z$ and $y_Z$ are determined by the qubits for $u = 1$ and 0 of the chain, 
respectively.
For simplicity, we represent the above two-qubit chain compactly by 
$(x_Z,y_Z,\alpha,\beta)$.
The formal correspondence between the different representations is summarized as
\begin{align}
&\mathrm{chain}(x_Z,y_Z,\alpha,\beta)
\nonumber \\
&= \Bigg[ \left( 
0, x_Z + \alpha, 2 - 2 \alpha + \beta , y_Z - 2 \left\lfloor \frac{y_Z}{2} \right\rfloor, \left\lfloor \frac{y_Z}{2}\right\rfloor 
\right) 
\nonumber \\
&\qquad 
\sim \left( 
1, y_Z + \alpha, 2 - 2 \alpha + \beta, x_Z - 2 \left\lfloor \frac{x_Z}{2} \right\rfloor, \left\lfloor \frac{x_Z}{2} \right\rfloor 
\right) \Bigg] 
\nonumber \\
&=[(0,2z_1+j_1+\alpha,2-2\alpha+\beta,j_0,z_0) 
\nonumber \\
&\qquad \sim (1,2z_0+j_0+\alpha,2-2\alpha+\beta,j_1,z_1)].
\label{eq:chain}
\end{align}
Examples of chains are shown in Fig.~\ref{fig:chains}.

\begin{figure}
\begin{center}
\includegraphics[width=0.8\columnwidth]{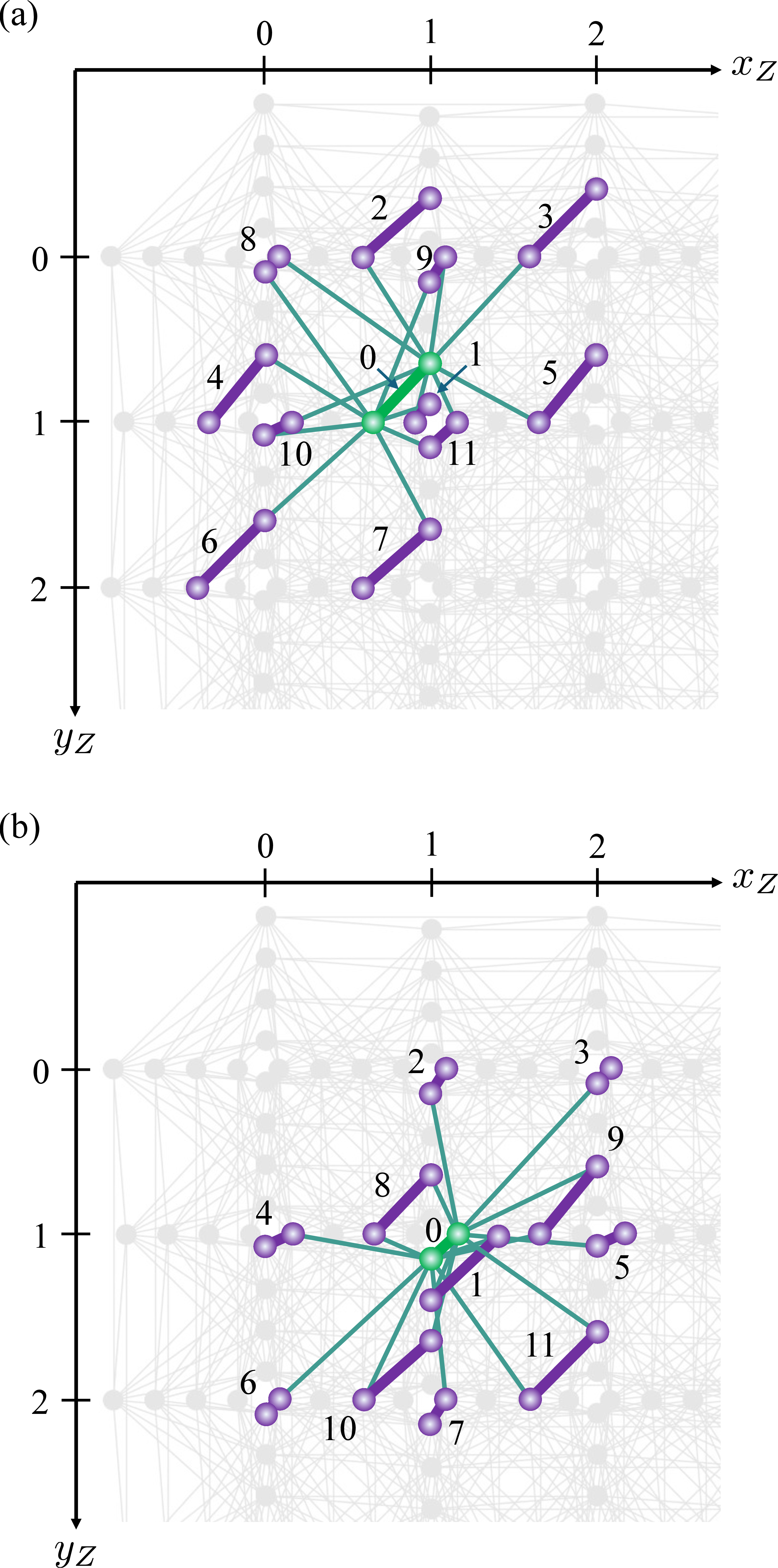}%
\caption{\label{fig:chain_couplings}
			Couplings to a chain, say, chain 0, used in our embedding in the Zephyr graph.
			The couplings of chains are based on physical couplers between qubits belonging to different chains.
			We only consider the internal couplers~\cite{K.Boothby2021}, 
			i.e., the couplers between physical qubits for $u=0$ and $1$ [Fig.~\ref{fig:Zephyr}].
			Chain 0 and chains which couple to it are labelled and highlighted.
			Edges representing the couplings to chain 0 via the internal couplers are also highlighted.
			Parameters of the labelled chains 
			when chain 0 is parameterized as $(x_Z, y_Z, \alpha, \beta)$ 
			are shown in Table~\ref{table:chain_couplings}.
			Two examples for parameters $(x_Z, y_Z, \alpha, \beta)$ of chain 0 are shown:
			(a)  $(1, 1, 0, 0)$ and (b) $(1, 1, 1, 0)$.
			In these examples, 
			all the fourth parameters of chains 2--11, 
			which are represented as $\beta'$ in Table~\ref{table:chain_couplings}, 
			are set to 0 $(= \beta)$, 
			but each of them can also be 1, 
			since whether the coupling between two chains exists does not depend on the value of their fourth parameters.}
\end{center}
\end{figure}

We specify how the introduced two-qubit chains couple to one another in the Zephyr graph.
These chain-to-chain couplings determine 
whether the two-body interactions required by the intermediate Hamiltonian $H'$ 
can be implemented on hardware.
As illustrated in Fig.~\ref{fig:chain_couplings}, 
the coupling of chains is based on physical couplers between qubits belonging to different chains.
Thus, coupling of chains means the existence of couplings between the physical qubits 
that constitute the two chains.
We only consider couplings of the two-qubit chains via the internal couplers~\cite{K.Boothby2021} of the Zephyr graph, 
i.e., the couplers between physical qubits for $u=0$ and $1$.
With this restriction, a chain $(x_Z,y_Z,\alpha,\beta)$ couples to the set of chains 
summarized in Table~\ref{table:chain_couplings} and depicted in Fig.~\ref{fig:chain_couplings}.
Some of these couplings may not exist near the boundary of the hardware graph.
Whether the coupling between two chains
exists does not depend on the value of $\beta$, except for the coupling to
$(x_Z,y_Z,\alpha,\bar{\beta})$.
Table~I also identifies which neighboring chains have two-qubit coupling,
meaning that both physical qubits in a chain couple to the reference chain
$(x_Z,y_Z,\alpha,\beta)$.
This will be discussed later.

\begin{table}
\caption{Chains coupled to chain $(x_Z,y_Z,\alpha,\beta)$. 
Some of the chains may not exist near the boundary of the hardware graph.
The second column shows labels of the corresponding chains in Fig.~\ref{fig:chain_couplings}. 
The third column,
two-qubit coupling (TQC), indicates if both the two qubits in the chain couple to 
the reference
chain $(x_Z,y_Z,\alpha,\beta)$. 
The fourth column shows parameter $u$ of qubits in the chain coupling to qubits in the reference chain.
Note that if TQC in the third column is true,
both the two qubits $u=0$ and 1 couple to the reference chain 
as shown in the fourth column.
Otherwise, only a single qubit couples to it.
$\bar{\alpha}=1-\alpha$ and $\bar{\beta}=1-\beta$. 
$\beta'$ can be 0 and 1.}
\label{table:chain_couplings}

\centering
\begin{tabular}{l c c c}
\hline
Coupled chain & Fig.~\ref{fig:chain_couplings} & TQC & coupled-qubits $u$ \\
\hline
$(x_Z, y_Z, \alpha, \bar{\beta})$ & 1 & True & 0, 1 \\
$(x_Z, y_Z-1, \alpha, \beta')$ & 2 & False & $\alpha$ \\
$(x_Z+1, y_Z-1, \alpha, \beta')$ & 3 & False & $\alpha$ \\
$(x_Z-1, y_Z, \alpha, \beta')$ & 4 & False & $\bar{\alpha}$ \\
$(x_Z+1, y_Z, \alpha, \beta')$ & 5 & False & $\alpha$ \\
$(x_Z-1, y_Z+1, \alpha, \beta')$ & 6 & False & $\bar{\alpha}$ \\
$(x_Z, y_Z+1, \alpha, \beta')$ & 7 & False & $\bar{\alpha}$ \\
$(x_Z-\bar{\alpha}, y_Z-\bar{\alpha}, \bar{\alpha}, \beta')$ & 8 & True & 0, 1\\
$(x_Z+\alpha, y_Z-\bar{\alpha}, \bar{\alpha}, \beta')$ & 9 & True & 0, 1\\
$(x_Z-\bar{\alpha}, y_Z+\alpha, \bar{\alpha}, \beta')$ & 10 & True & 0, 1\\
$(x_Z+\alpha, y_Z+\alpha, \bar{\alpha}, \beta')$ & 11 & True & 0, 1\\
\hline
\end{tabular}
\end{table}



\subsection{Embedding}
\label{sec:embedding}

To describe our embedding, 
we first illustrate the mapping $\phi$ of a small set of spins as explicit examples, 
and then introduce operators that allow us to determine the chains for arbitrary spins by iteration.
As a starting point, consider a pair consisting of a parity spin and the corresponding
ancillary spin at the same coordinate $(x_P,y_P)$.
We map $(x_P,y_P,0)$ and $(x_P,y_P,1)$ to two chains that share the same
$(x_Z,y_Z,\alpha)$ and differ only in $\beta$:
\begin{equation}
\phi(x_P,y_P,0) = (x_Z,y_Z,\alpha,\beta),
\label{eq:phi_parity_0}
\end{equation}
\begin{equation}
\phi(x_P,y_P,1) = (x_Z,y_Z,\alpha,\bar{\beta}).
\label{eq:phi_ancillary_0}
\end{equation}
For simplicity, 
we explain the embedding where every parity and ancillary spin are mapped to the chains for $\beta$ and $\bar{\beta}$, respectively.
However, since whether the coupling between two chains exists does not depend on the value of $\beta$, 
except for the special pairing $(x_Z,y_Z,\alpha,\bar{\beta})$, 
one may also swap the roles of $\beta$ and $\bar{\beta}$ in this assignment without changing the feasibility of couplings.
From the initial pair, we specify how neighbors of the pair are mapped as follows:
\begin{equation}
\phi(x_P+1,y_P,0) = (x_Z+\alpha,y_Z-\bar{\alpha},\bar{\alpha},\beta),
\label{eq:phi_parity_1}
\end{equation}
\begin{equation}
\phi(x_P+1,y_P,1) = (x_Z+1,y_Z,\alpha,\bar{\beta}),
\label{eq:phi_ancillary_1}
\end{equation}
\begin{equation}
\phi(x_P+1,y_P-1,0) = (x_Z+1,y_Z-1,\alpha,\beta),
\label{eq:phi_parity_2}
\end{equation}
\begin{equation}
\phi(x_P+1,y_P-1,1) = (x_Z+1,y_Z-1,\alpha,\bar{\beta}),
\end{equation}
\begin{equation}
\phi(x_P+1,y_P+1,0) = (x_Z+\alpha,y_Z+\alpha,\bar{\alpha},\beta),
\end{equation}
\begin{equation}
\phi(x_P+1,y_P+1,1) = (x_Z+\alpha,y_Z+\alpha,\bar{\alpha},\bar{\beta}).
\label{eq:phi_ancillary_3}
\end{equation}
These mappings are determined so that 
the required couplings in the intermediate Hamiltonian can be realized 
using the couplers available between the corresponding chains in the Zephyr graph, 
which will be verified later.
Once these local rules are fixed, the mapping of further spins can be determined iteratively.


To describe the iteration compactly, we introduce translation operators acting on chain labels:
\begin{equation}
T_{+0}(x_Z,y_Z,\alpha,\beta) = (x_Z+\alpha,y_Z-\bar{\alpha},\bar{\alpha},\beta),
\end{equation}
\begin{equation}
T_{+-}(x_Z,y_Z,\alpha,\beta) = (x_Z+1,y_Z-1,\alpha,\beta),
\end{equation}
\begin{equation}
T_{++}(x_Z,y_Z,\alpha,\beta) = (x_Z+\alpha,y_Z+\alpha,\bar{\alpha},\beta),
\end{equation}
which give the chains to which neighboring parity spins are mapped. 
When these operators act on the chain assigned to a parity spin
$(x_P,y_P,0)$, they reproduce the neighbor relations written above:
\begin{equation}
T_{+0}\phi(x_P,y_P,0) = \phi(x_P+1,y_P,0),
\label{eq:T_+0}
\end{equation}
\begin{equation}
T_{+-}\phi(x_P,y_P,0) = \phi(x_P+1,y_P-1,0),
\label{eq:T_+-}
\end{equation}
\begin{equation}
T_{++}\phi(x_P,y_P,0) = \phi(x_P+1,y_P+1,0).
\label{eq:T_++}
\end{equation}
In the iteration, we use multiple products of the last two translations:
\begin{equation}
T^{c}_{+-}(x_Z,y_Z,\alpha,\beta) = (x_Z+c,y_Z-c,\alpha,\beta),
\end{equation}
\begin{align}
&T^{c}_{++}(x_Z,y_Z,\alpha,\beta)
\nonumber \\
&= (x_Z+f(c,\alpha),y_Z+f(c,\alpha),\alpha+g(c,\alpha),\beta),
\end{align}
where
\begin{equation}
f(c,\alpha) = \left\lfloor \frac{c}{2}\right\rfloor + 2\alpha\left(\frac{c}{2}-\left\lfloor \frac{c}{2}\right\rfloor\right),
\end{equation}
\begin{equation}
g(c,\alpha) = 2(1-2\alpha)\left(\frac{c}{2}-\left\lfloor \frac{c}{2}\right\rfloor\right)
\end{equation}
for $c=0,\pm1,\pm2,\ldots$.
We also use 
\begin{equation}
T^{-1}_{+0}(x_Z,y_Z,\alpha,\beta)
= (x_Z - \alphabar, y_Z + \alpha ,\alphabar, \beta)
\end{equation}
and the commutativity of the operators $T_{+0}$, $T_{+-}$, and $T_{++}$, e.g.,   
\begin{align}
T_{+-}T_{++} (x_Z,y_Z,\alpha,\beta)
&= T_{++}T_{+-} (x_Z,y_Z,\alpha,\beta)
\nonumber \\
&= (x_Z + 1 + \alpha, y_Z - 1 + \alpha, \alphabar, \beta).
\label{eq:commutative_T}
\end{align}

To obtain the chain for an ancillary
spin from that of its associated parity spin, 
we introduce two operators:
\begin{equation}
A(x_Z,y_Z,\alpha,\beta) = (x_Z,y_Z,\alpha,\bar{\beta}),
\label{eq:A}
\end{equation}
\begin{align}
A'(x_Z,y_Z,\alpha,\beta)
&=AT_{++}(x_Z,y_Z,\alpha,\beta)
\nonumber \\
&=(x_Z+\alpha,y_Z+\alpha,\bar{\alpha},\bar{\beta}).
\label{eq:A'}
\end{align}
The first one $A$ reproduces the relation of the pair in Eqs.~(\ref{eq:phi_parity_0}) and (\ref{eq:phi_ancillary_0}) as
\begin{equation}
\phi(x_P,y_P,1) = A \phi(x_P,y_P,0).
\label{eq:Aphi}
\end{equation}
This can be also applied to the pairs in Eqs.~(\ref{eq:phi_parity_2})--(\ref{eq:phi_ancillary_3}) as
\begin{align}
\phi(x_P + 1,y_P - 1,1) 
&= A \phi(x_P + 1, y_P - 1, 0),
\\
\phi(x_P + 1, y_P + 1, 1) 
&= A \phi(x_P + 1, y_P + 1, 0).
\end{align}
These relations with Eqs.~(\ref{eq:T_+-}) and (\ref{eq:T_++}) show that 
$T_{+-}$ and $T_{++}$ keep that 
an ancillary-spin chain is obtained with $A$ acting on the corresponding parity-spin chain.
However, 
$T_{+0}$ causes the required operator to switch from $A$ to $A'$.
In fact $A'$ reproduces the relation of the pair in Eqs.~(\ref{eq:phi_parity_1}) and (\ref{eq:phi_ancillary_1}) as 
\begin{equation}
\phi(x_P+1,y_P,1)
= A' \phi(x_P+1,y_P,0),
\end{equation}
and this parity-spin chain is obtained with $T_{+0}$ acting on the first parity-spin chain 
as in Eq.~(\ref{eq:T_+0}).

Using the above operators and the chain for the first parity spin [Eq.~(\ref{eq:phi_parity_0})] 
whose associated ancillary spin is mapped as Eq.~(\ref{eq:phi_ancillary_0}),
the chain for another spin $(x'_P,y'_P,a)$ is given by
\begin{align}
&\phi(x'_P,y'_P,a)
\nonumber \\
&=
A'^{c_{+0}a}
A^{(1-c_{+0})a}
T^{c_{+0}}_{+0}
T^{c_{+-}}_{+-}
T^{c_{++}}_{++}
\phi(x_P,y_P,0)
\label{eq:mapping}
\end{align}
with exponents determined from the coordinate differences 
in the intermediate Hamiltonian as follows:
\begin{align}
c_{+0} &= 0 \text{ or } 1, 
\\
c_{+0} &\equiv \Delta_+(x'_P, x_P, y'_P, y_P) \pmod 2, 
\\
c_{+-}&= \left[ \Delta_-(x'_P, x_P, y'_P, y_P)-c_{+0} \right]/2, 
\\
c_{++}&= \left[\Delta_+(x'_P, x_P, y'_P, y_P)-c_{+0} \right]/2, 
\label{eq:c_++}
\end{align}
where
\begin{equation}
\Delta_\pm(x'_P, x_P, y'_P, y_P)
= (x'_P-x_P) \pm (y'_P-y_P).
\end{equation}
We can also obtain the chain for another spin $(x''_P,y''_P,a)$ not only from the first chain 
but also from another parity-spin chain given by Eq.~(\ref{eq:mapping}) for $a = c_{+0} = 0$ as
\begin{align}
\phi(x''_P,y''_P,a)
=&
A'^{c'_{+0}a}
A^{(1-c'_{+0})a}
T^{c'_{+0}}_{+0}
T^{c'_{+-}}_{+-}
T^{c'_{++}}_{++}
\phi(x_P,y_P,0)
\nonumber \\
=&
A'^{c'_{+0}a}
A^{(1-c'_{+0})a}
T^{c'_{+0}}_{+0}
\nonumber \\
& \times T^{(c'_{+-} - c_{+-})}_{+-}
T^{(c'_{++} - c_{++})}_{++}
\phi(x'_P,y'_P,0), 
\label{eq:mapping_translation}
\end{align}
where
\begin{align}
c'_{+0} 
&= 0 \text{ or } 1, 
\\
c'_{+0} 
&\equiv \Delta_+(x''_P, x_P, y''_P, y_P) \pmod 2
\nonumber \\
&\equiv \Delta_+(x''_P, x'_P, y''_P, y'_P) \pmod 2, 
\end{align}
\begin{align}
c'_{+-}
&=\left[ \Delta_-(x''_P, x_P, y''_P, y_P) - c'_{+0} \right]/2, 
\nonumber \\
&=\left[ \Delta_-(x''_P, x'_P, y''_P, y'_P) -c'_{+0} \right]/2 + c_{+-}, 
\end{align}
\begin{align}
c'_{++}
&=\left[ \Delta_+(x''_P, x_P, y''_P, y_P) - c'_{+0} \right] /2
\nonumber \\
&=\left[ \Delta_+(x''_P, x'_P, y''_P, y'_P) - c'_{+0} \right]/2 + c_{++}
\end{align}
We have used the commutativity of $T_{+-}$ and $T_{++}$ [Eq.~(\ref{eq:commutative_T})].
Equations~(\ref{eq:mapping}) and (\ref{eq:mapping_translation}) have assumed that 
the chains for the reference pair of parity and ancillary spins are related as in Eqs~(\ref{eq:phi_parity_0}) and (\ref{eq:phi_ancillary_0})
and also characterized by operator $A$ [Eq.~(\ref{eq:A})].
However, the other type of pairs, 
say, $(\tilde{x}_P,\tilde{y}_P,0)$ and $(\tilde{x}_P,\tilde{y}_P,1)$, 
characterized by $A'$ [Eq.~(\ref{eq:A'})]
can be also used as a reference to obtain the chain for a different spin as 
\begin{align}
&\phi(x'_P,y'_P,a)
\nonumber \\
&=
A'^{c_{+0}a}
A^{(1-c_{+0})a}
T^{c_{+0}}_{+0}
T^{c_{+-}}_{+-}
T^{c_{++}}_{++}
T^{-1}_{+0}
\phi(\tilde{x}_P,\tilde{y}_P,0)
\nonumber \\
&=
A'^{(\tilde{c}_{+0}+1)a}
A^{-\tilde{c}_{+0}a}
T^{\tilde{c}_{+0}}_{+0}
T^{\tilde{c}_{+-}}_{+-}
T^{\tilde{c}_{++}}_{++}
\phi(\tilde{x}_P,\tilde{y}_P,0), 
\label{eq:mapping_2}
\end{align}
where
\begin{align}
\tilde{c}_{+0} &= -1 \text{ or } 0, 
\\
\tilde{c}_{+0} &\equiv \Delta_+(x'_P, \tilde{x}_P, y'_P, \tilde{y}_P) \pmod 2, 
\\
\tilde{c}_{+-} &=\left[ \Delta_-(x'_P, \tilde{x}_P, y'_P, \tilde{y}_P \right] -\tilde{c}_{+0})/2, 
\\
\tilde{c}_{++} &= \left[ \Delta_+(x'_P, \tilde{x}_P, y'_P, \tilde{y}_P) -\tilde{c}_{+0} \right]/2.
\end{align}



\begin{table}
\caption{The spins coupled to spin $(x_P,y_P,0)$ in the intermediate Hamiltonian $H'$ [Eq.~(7)] 
and the chains in the Zephyr graph assigned to the spins as Eq.~(\ref{eq:mapping}), 
provided that spins $(x_P,y_P,0)$ and $(x_P,y_P,1)$ are mapped to chains $(x_Z,y_Z,\alpha,\beta)$ and $(x_Z,y_Z,\alpha,\betabar)$, 
respectively.
The third and fourth columns show labels of the corresponding chains 
in Table~\ref{table:chain_couplings} (Fig.~\ref{fig:chain_couplings}) and Fig.~\ref{fig:mapping}, respectively. 
$\bar{\alpha}=1-\alpha$ and $\bar{\beta}=1-\beta$.}
\label{table:correspondence}
\centering
\begin{tabular}{l l c c}
\hline
Coupled spins & Assigned chains & Table~\ref{table:chain_couplings} & Fig.~\ref{fig:mapping} \\
\hline
$(x_P - 1, y_P - 1, 0)$ & $(x_Z - \bar{\alpha}, y_Z - \bar{\alpha}, \bar{\alpha}, \beta)$ & 8 & 1 \\
$(x_P, y_P - 1, 0)$ & $(x_Z, y_Z - 1, \alpha, \beta)$ & 2 & 2 \\
$(x_P + 1, y_P - 1, 0)$ & $(x_Z + 1, y_Z - 1, \alpha, \beta)$ & 3 & 3 \\
$(x_P - 1, y_P, 0)$ & $(x_Z - 1, y_Z, \alpha, \beta)$ & 4 & 4 \\
$(x_P + 1, y_P, 0)$ & $(x_Z + \alpha, y_Z - \bar{\alpha}, \bar{\alpha}, \beta)$ & 9 & 5 \\
$(x_P - 1, y_P + 1, 0)$ & $(x_Z - 1, y_Z + 1, \alpha, \beta)$ & 6 & 6 \\
$(x_P, y_P + 1, 0)$ & $(x_Z - \bar{\alpha}, y_Z + \alpha, \bar{\alpha}, \beta)$ & 10 & 7 \\
$(x_P + 1, y_P + 1, 0)$ & $(x_Z + \alpha, y_Z + \alpha, \bar{\alpha}, \beta)$ & 11 & 8 \\
$(x_P - 1, y_P - 1, 1)$ & $(x_Z - \bar{\alpha}, y_Z - \bar{\alpha}, \bar{\alpha}, \bar{\beta})$ & 8 & 9 \\
$(x_P, y_P - 1, 1)$ & $(x_Z + \alpha, y_Z - \bar{\alpha}, \bar{\alpha}, \bar{\beta})$ & 9 & 10 \\
$(x_P - 1, y_P, 1)$ & $(x_Z - \bar{\alpha}, y_Z + \alpha, \bar{\alpha}, \bar{\beta})$ & 10 & 11 \\
$(x_P, y_P, 1)$ & $(x_Z, y_Z, \alpha, \bar{\beta})$ & 1 & 12 \\
\hline
\end{tabular}
\end{table}

\begin{figure*}[t]
\includegraphics[width=2.0\columnwidth]{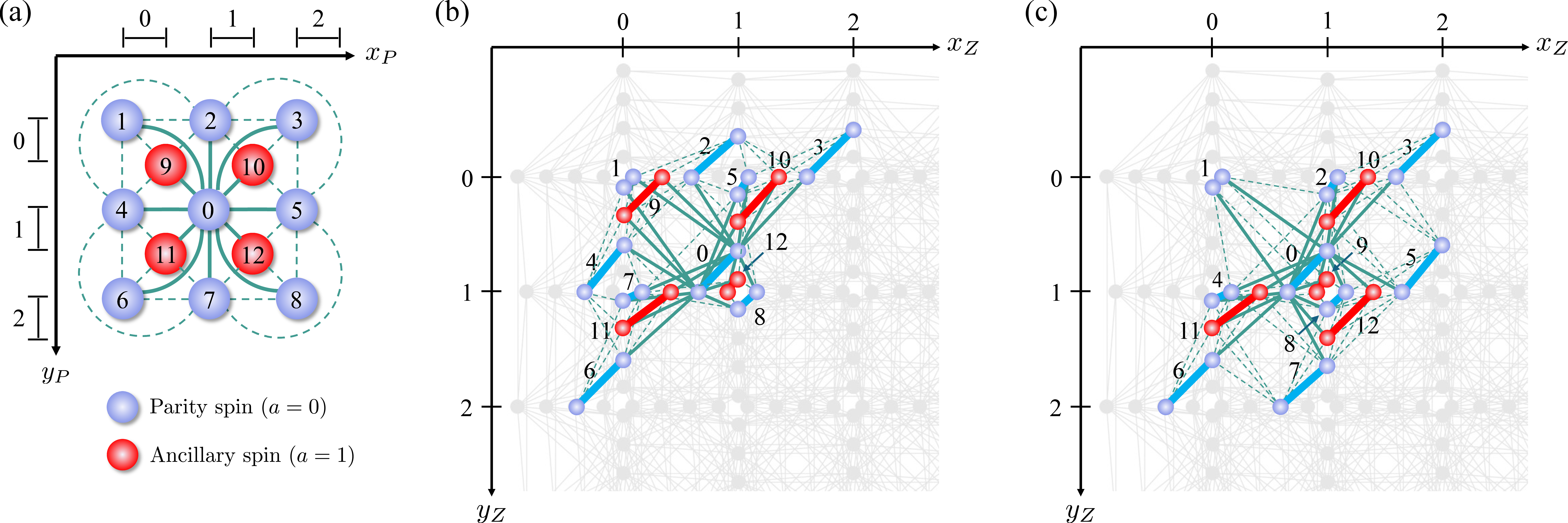}%
\caption{Mapping of a spin and its coupled spins in the intermediate Hamiltonian [Eq.~(\ref{eq:H_intermediate})]
to two-qubit chains in the Zephyr graph.
(a) Lattice for the intermediate Hamiltonian for $3\times 3$ parity spins and $2\times 2$ ancillary spins.
Edges for couplings to spin 0 are depicted as solid lines, 
while those for the other couplings are depicted as dashed lines.
See Fig.~\ref{fig:LHZ-intermediate_lattice}(b) for the coordinated description.
The spins are labelled. 
For example, spin $(1, 1, 0)$ is labelled as 0.
(b), (c) Chains in the Zephyr graph assigned to the spins in (a) 
as (b) Eq.~(\ref{eq:mapping}) and (c) Eq.~(\ref{eq:mapping_2}), 
provided that
spins 0 and 12 are mapped to 
(b) chains $(1,1,0,0)$ and $(1,1,0,1)$ and (c) chains $(1,1,0,0)$ and $(1,1,1,1)$, 
respectively.
See, Fig.~\ref{fig:chains} for the coordinate description.
The chains have the same labels as their corresponding spins in (a).
Parameters of the labelled chains 
when chain 0 is parameterized as $(x_Z, y_Z, \alpha, \beta)$ 
are shown in (b) Table~\ref{table:correspondence} and (c) Table~\ref{table:correspondence_2}.
}
\label{fig:mapping}
\end{figure*}

\begin{table}
\caption{The spins coupled to spin $(x_P,y_P,0)$ in the intermediate Hamiltonian $H'$ [Eq.~(7)] 
and the chains in the Zephyr graph assigned to the spins as Eq.~(\ref{eq:mapping_2}), 
provided that spins $(\tilde{x}_P,\tilde{y}_P,0)$ and $(\tilde{x}_P,\tilde{y}_P,1)$ are mapped to chains $(x_Z,y_Z,\alpha,\beta)$ and $(x_Z+1,y_Z,\alpha,\bar{\beta})$, 
respectively.
The third and fourth columns show labels of the corresponding chains 
in Table~\ref{table:chain_couplings} and Fig.~\ref{fig:mapping}, respectively. 
$\bar{\alpha}=1-\alpha$ and $\bar{\beta}=1-\beta$.}
\label{table:correspondence_2}
\centering
\begin{tabular}{l l c c}
\hline
Coupled spins & Assigned chains & Table~\ref{table:chain_couplings} & Fig.~\ref{fig:mapping} \\
\hline
$(\tilde{x}_P - 1, \tilde{y}_P - 1, 0)$ & $(x_Z - \bar{\alpha}, y_Z - \bar{\alpha}, \bar{\alpha}, \beta)$ & 8 & 1 \\
$(\tilde{x}_P, \tilde{y}_P - 1, 0)$ & $(x_Z + \alpha, y_Z - \alphabar, \alphabar, \beta)$ & 2 & 2 \\
$(\tilde{x}_P + 1, \tilde{y}_P - 1, 0)$ & $(x_Z + 1, y_Z - 1, \alpha, \beta)$ & 3 & 3 \\
$(\tilde{x}_P - 1, \tilde{y}_P, 0)$ & $(x_Z - \alphabar, y_Z + \alpha, \alphabar, \beta)$ & 10 & 4 \\
$(\tilde{x}_P + 1, \tilde{y}_P, 0)$ & $(x_Z + 1, y_Z, \alpha, \beta)$ & 5 & 5 \\
$(\tilde{x}_P - 1, \tilde{y}_P + 1, 0)$ & $(x_Z - 1, y_Z + 1, \alpha, \beta)$ & 6 & 6 \\
$(\tilde{x}_P, \tilde{y}_P + 1, 0)$ & $(x_Z, y_Z + 1, \alpha, \beta)$ & 7 & 7 \\
$(\tilde{x}_P + 1, \tilde{y}_P + 1, 0)$ & $(x_Z + \alpha, y_Z + \alpha, \bar{\alpha}, \beta)$ & 11 & 8 \\
$(\tilde{x}_P - 1, \tilde{y}_P - 1, 1)$ & $(x_Z, y_Z, \alpha, \bar{\beta})$ & 1 & 9 \\
$(\tilde{x}_P, \tilde{y}_P - 1, 1)$ & $(x_Z + \alpha, y_Z - \bar{\alpha}, \bar{\alpha}, \bar{\beta})$ & 9 & 10 \\
$(\tilde{x}_P - 1, \tilde{y}_P, 1)$ & $(x_Z - \bar{\alpha}, y_Z + \alpha, \bar{\alpha}, \bar{\beta})$ & 10 & 11 \\
$(\tilde{x}_P, \tilde{y}_P, 1)$ & $(x_Z + \alpha, y_Z + \alpha, \alphabar, \bar{\beta})$ & 11 & 12 \\
\hline
\end{tabular}
\end{table}

To confirm the validity of our embedding, 
consider the parity spin $(x_P,y_P,0)$.
In the intermediate Hamiltonian $H'$ [Eq.~(\ref{eq:H_intermediate})], 
this spin couples to the spins listed in the column
``coupled spins'' of Table~\ref{table:correspondence}.
In our embedding based on Eq.~(\ref{eq:mapping}), 
each of those coupled spins is mapped to the corresponding chain 
listed in the column ``assigned chains'' of the same table, 
given that 
spins $(x_P,y_P,0)$ and $(x_P,y_P,1)$ are mapped 
to chains $(x_Z,y_Z,\alpha,\beta)$ and $(x_Z,y_Z,\alpha,\betabar)$, 
respectively. 
Importantly, all chains appearing in the ``assigned chains'' column are among the chains
listed in Table~\ref{table:chain_couplings} as being coupled to $(x_Z,y_Z,\alpha,\beta)$.
This means that there exist physical couplers between the chain assigned to $(x_P,y_P,0)$
and the chains assigned to every spin that couples to it in $H'$.
The mapping of spin $(x_P,y_P,0)$ and the coupled spins is shown in Fig.~\ref{fig:mapping}(a) and (b).
Table~\ref{table:correspondence_2} and Fig.~\ref{fig:mapping}(c) show the corresponding fact
for the case where spins $(x_P,y_P,0)$ and $(x_P,y_P,1)$ are mapped 
to chains $(x_Z,y_Z,\alpha,\beta)$ and $(x_Z+\alpha,y_Z+\alpha,\alphabar,\betabar)$, 
respectively. 
This pair is characterized by $A'$ [Eq.~\ref{eq:A'}].
This argument extends to all parity spins owing to the translational symmetry of the construction [Eq.~(\ref{eq:mapping_translation})].
It also guarantees implementability of couplings associated with ancillary spins, 
since ancillary spins couple only to parity spins in $H'$, 
and all couplings associated with parity spins can be implemented as shown above.
Therefore, all couplings of spins in $H'$ can be implemented as couplings between physical
qubits belonging to the corresponding chains in the Zephyr graph under our embedding.

We now discuss the effects of boundary conditions. 
When spins are assigned to chains according to the embedding rules given by Eqs.~(\ref{eq:mapping}) and (\ref{eq:mapping_2}), 
the embedding remains feasible 
even if a chain is located at the boundary of the Zephyr graph, 
provided that the chain itself exists on the graph, 
that is, 
the chain to be assigned does not extend beyond the graph.
This property follows from the fact that, 
in the Zephyr graph, 
the existence of a coupler between two qubits is determined solely by their relative positions 
and is independent of their absolute location within the graph. 
For example, 
an internal coupler~\cite{K.Boothby2021} connects the two qubits $(0, 2w + 1 - \alpha, k, j, z - j\beta)$ and $(1, 2z + 1 - \beta, h, i, w - i\alpha)$, 
where $\alpha, \beta=0$ or 1.
Aside from hardware defects, 
if two qubits satisfy this positional relationship, 
the corresponding internal coupler is always present. 
In other words, 
there are no cases in which two qubits with the required relative configuration exist 
while the associated internal coupler is absent.
Therefore, 
the only boundary effect is whether a chain extends beyond the physical limits of the graph. 
For example, 
in the square lattice shown in Fig.~\ref{fig:mapping}(a), 
it is sufficient to ensure that 
the parity spins located 
at the upper-right corner ($x_P$ maximal and $y_P$ minimal) and the lower-left corner ($x_P$ minimal and $y_P$ maximal) 
can be embedded; 
all other spins can then be embedded automatically.

The above discussion has focused on edge effects in the Zephyr graph. 
We now consider the effects of boundary spins 
in the intermediate-Hamiltonian graph. 
For parity spins located at the boundary of the intermediate-Hamiltonian graph, 
the corresponding embedding chains can, in some cases, 
be reduced to a single qubit. 
This property allows the construction of a more efficient embedding variant. 
Further details are presented in Appendix.



\subsection{Optionality}
\label{sec:optionality}

We next discuss the optionality of our embedding scheme, 
which can be useful to avoid inactive qubits.
In our embedding, the chains assigned to the ancillary spins that couple to the parity spin $(x_P, y_P, 0)$,
which are the last four entries in Tables~\ref{table:correspondence} and \ref{table:correspondence_2}, 
have two-qubit coupling as indicated in Table~\ref{table:chain_couplings}.
This means that, for each of those ancillary-spin chains, both physical qubits in the chain
couple to the chains for the neighboring parity spins.
Thus, 
each ancillary-spin chain can be reduced to either qubit in the chain without losing the
necessary couplings to neighboring parity spins.
In other words, when realizing an ancillary spin on hardware, 
we can choose either of the two physical qubits.
The choice between the two qubits does not affect the embedding of any other spins.
Therefore, 
provided that the physical qubits have identical properties, 
there is no intrinsic preference for either qubit.
Note that
all ancillary-spin chains have this optionality 
regardless of the size of the problem being embedded, 
since our embedding is constructed with the translation symmetry [Eq.~(\ref{eq:mapping_translation})].

\begin{figure*}[t]
\includegraphics[width=2.0\columnwidth]{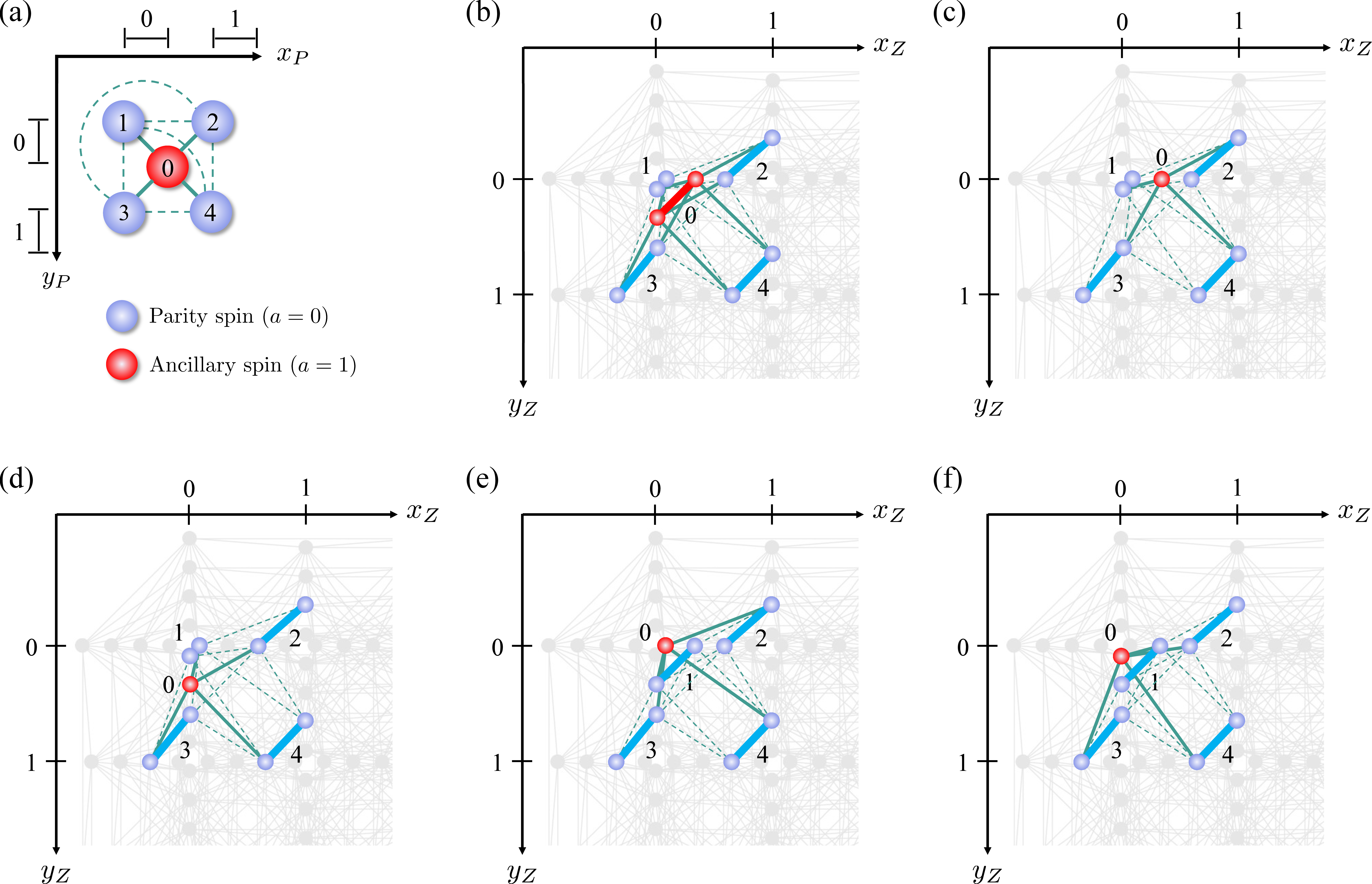}%
\caption{Options of qubits to be used for ancillary spins.
(a) Lattice for the intermediate Hamiltonian $H'$ [Eq~(\ref{eq:H_intermediate})] for $2\times 2$ parity spins and an ancillary spin.
Edges for couplings to the ancillary spin are depicted as solid lines, 
while those for the other couplings between parity spins are depicted as dashed lines.
See Fig.~\ref{fig:LHZ-intermediate_lattice}(b) for the coordinated description.
(b) Our embedding of $H'$ in the Zephyr graph based on Eq.~(\ref{eq:mapping}), 
provided that 
spin 1 and 0 are mapped to chains $(0, 0, 1, 0)$ and $(0, 0, 1, 1)$, 
respectively.
See, Fig.~\ref{fig:chains} for the coordinate description.
(c), (d) Variants of the embedding in (a).
The chain for the ancillary spin is reduced to either qubit.
Both the selected qubits couple to all chains for neighboring parity spins, 
meaning that
we can avoid an inactive qubit 
if one of the two qubits in the ancillary-spin chain is unavailable.
(e), (f) Extra variants of the embedding in (a).
$\beta$ is swapped between parity- and ancillary-spin chains 
which have the same $x_Z$, $y_Z$, and $\alpha$.
The updated ancillary-spin chain can be reduced to either qubit as in (c) and (d), 
meaning that
we can also avoid a possible inactive qubit in a parity-spin chain in the first embedding.
}
\label{fig:ancillary_options}
\end{figure*}

This optionality is especially useful in realistic hardware settings where some qubits may be
inactive.
If one of the two qubits in an ancillary-spin chain is unavailable, 
we can simply select the other qubit in the chain, 
thereby preserving the embedding while avoiding the inactive qubit 
as illustrated in Fig.~\ref{fig:ancillary_options}(a)--(d).
We can also extend this optionality to avoid inactive qubits in parity-spin chains.
If one of the two qubits in a parity-spin chain is unavailable, 
we swap $\beta$ with the ancillary-spin chain sharing the same $x_Z, y_Z$, and $\alpha$, 
and select the active qubit for the ancillary spin as illustrated in Fig.~\ref{fig:ancillary_options}(a), (e), and (f). 
The embedding after this swap still works well, 
since two-qubit coupling and
whether the coupling between two chains exists do not depend on $\beta$ 
[Table~\ref{table:chain_couplings}].

Note that 
the ability to reduce each chain for ancillary spins to a single qubit reduces the number of required physical qubits per
parity spin in the parity Hamiltonian [Eq.\ (1)] to three, 
since the number of ancillary spins is roughly equal to the number of parity spins. 
This is less than the number of qubits required in the previously proposed method to embed the
parity Hamiltonian in the Pegasus graph~\cite{M.Cattelan2025}.



We can also avoid inactive qubits in the Zephyr graph 
by leveraging options in the parity Hamiltonian itself.
Specifically, 
there could exist multiple options 
corresponding to different but logically equivalent choices in how the logical variables and their
interactions are encoded into the parity-spin representation.
If one parity-Hamiltonian representation leads to an embedding that intersects inactive qubits, we can instead
select an alternative parity-Hamiltonian representation of the same logical problem.
Because that alternative parity Hamiltonian maps to a different intermediate Hamiltonian, 
it can be embedded using different physical qubits, 
potentially avoiding the inactive ones.

\begin{figure*}[t]
\begin{center}
\includegraphics[width=1.8\columnwidth]{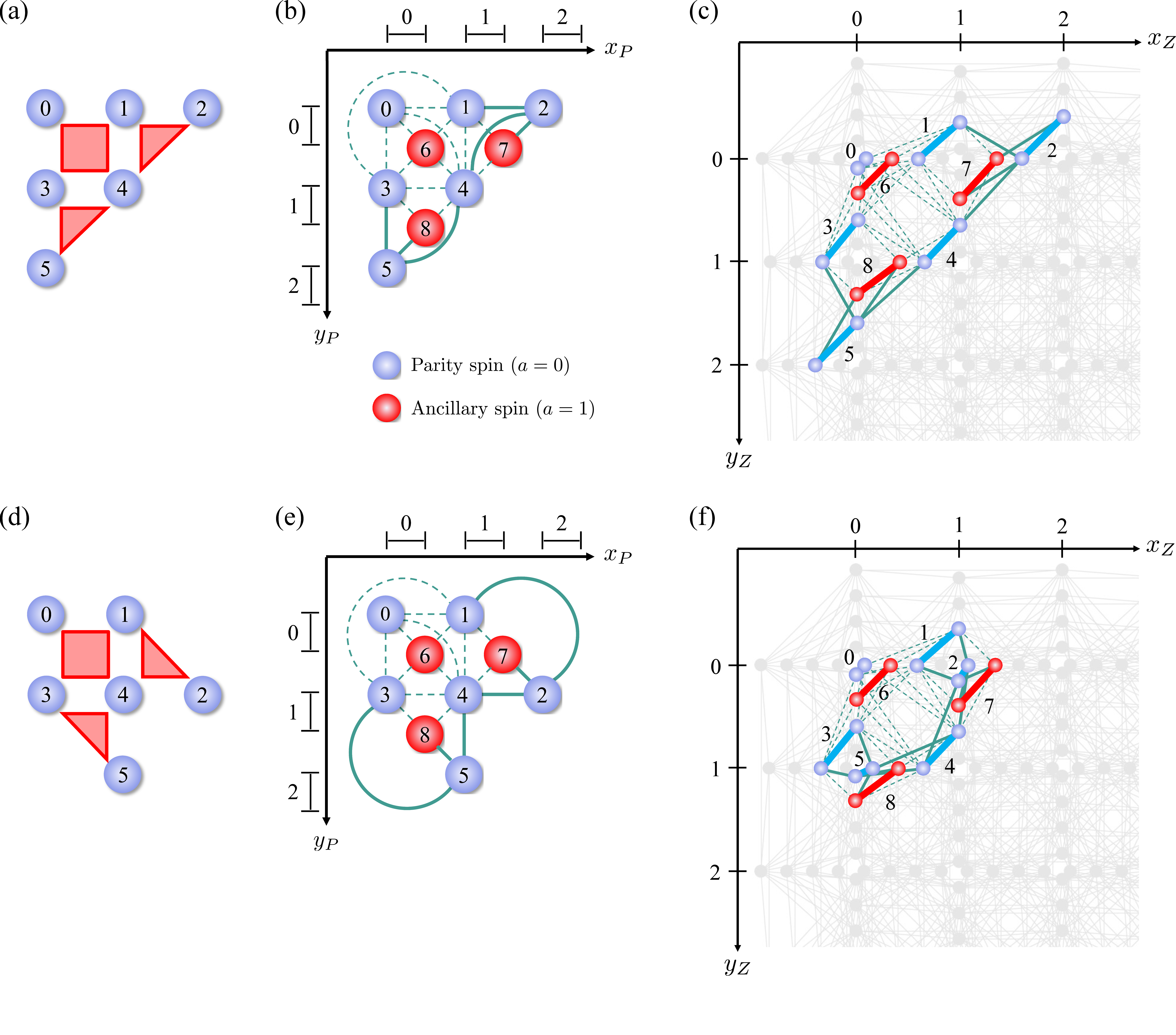}%
\caption{Changing the parity Hamiltonian to avoid inactive physical qubits in the Zephyr graph when embedding.
The logical Hamiltonian is considered for four Ising spins with the all-to-all two-body couplings.
(a) Lattice for the Parity Hamiltonian $H$ [Eq~(\ref{eq:H_parity})] for six parity spins which encodes the logical Hamiltonian.
The parity Hamiltonian includes three-body interactions between spins located at the vertices of triangles.
(b) Lattice for the intermediate Hamiltonian $H'$ [Eq~(\ref{eq:H_intermediate})] for six parity spins and three ancillary spins 
derived from the Parity Hamiltonian for (a).
Spins 2 and 5 are focused on, and edges representing the couplings to these spins are depicted as solid lines, 
while those for the other couplings are depicted as dashed lines.
See Fig.~\ref{fig:LHZ-intermediate_lattice}(b) for the coordinated description.
(c) Embedding the intermediate Hamiltonian for (b) in the Zephyr graph as Eq.~(\ref{eq:mapping}), 
provided that
spins 0 and 6 are mapped to chains $(0,0,1,0)$ and $(0,0,1,1)$, 
respectively.
See, Fig.~\ref{fig:chains} for the coordinate description.
(d) Lattice for the changed parity Hamiltonian from (a).
(e) Lattice for the intermediate Hamiltonian for (d).
(f) Embedding the intermediate Hamiltonian for (e), 
which uses qubits that are closer than in (c).
}
\label{fig:change_parity_Hamiltonian}
\end{center}
\end{figure*}

An example illustrating this idea is shown in Fig.~\ref{fig:change_parity_Hamiltonian}
for the case of a logical Hamiltonian for four spins with all-to-all two-body couplings.
A lattice for the parity Hamiltonian for this case is shown in Fig.~\ref{fig:change_parity_Hamiltonian}(a).
Spins in the intermediate Hamiltonian [Fig.~\ref{fig:change_parity_Hamiltonian}(b)] 
derived from the parity Hamiltonian are mapped to chains in the Zephyr graph 
using Eq.~(\ref{eq:mapping}), 
provided that
spins 0 and 6 are mapped to chains $(0,0,1,0)$ and $(0,0,1,1)$, 
respectively [Fig.~\ref{fig:change_parity_Hamiltonian}(c)].
The lattice for the parity Hamiltonian for the same logical Hamiltonian is also represented as in [Fig.~\ref{fig:change_parity_Hamiltonian}(d)].
The embedding for this case  [Fig.~\ref{fig:change_parity_Hamiltonian}(f)] uses qubits 
that are closer to each other than in the former case.
Consequently,
the second parity Hamiltonian is embedded in the Zephyr graph for $m = 1$, 
while the first one needs $m = 2$.  

\section{Efficiency of the proposed embedding scheme}
\label{sec:efficiency}

\subsection{Number of physical qubits}

To evaluate the efficiency of the proposed embedding scheme, 
we first compare the number of physical qubits used by our embeddings 
with that obtained using \texttt{find\_embedding}, 
a known heuristic minor-embedding algorithm~\cite{J.Cai2014arXiv}. 
We will discuss computational costs in the next subsection.
For this comparison, 
we consider intermediate Hamiltonians transformed from parity Hamiltonians 
defined on square lattices with various side lengths \(L\). 
For example, 
the parity Hamiltonian illustrated 
in Fig.~~\ref{fig:LHZ-intermediate_lattice}(a) corresponds to \(L=3\). 
Although the efficiency of our scheme compared with the heuristic might depend on the global shape of the lattice for the parity Hamiltonian, 
we focus on the square case here, 
and the investigation of other shapes is left for future work.

The total number \(N_{1}\) of physical qubits used in our embedding is determined uniquely as 
\begin{equation} 
N_{1} = 2L^{2}+(L-1)^{2}. 
\label{eq:N1} 
\end{equation} 
Here, the first and second terms represent the physical qubits 
used for the parity and ancillary spins, respectively. 
We also consider the variant of the proposed embedding 
described in Appendix. 
In this variant, 
some of the parity-spin chains located at the boundaries of the lattice are reduced 
from two-qubit chains to single physical qubits. 
As shown in Appendix, 
the number of parity-spin chains 
that can be reduced in this manner is \(2L-4\). 
The resulting number \(N_{2}\) of physical qubits is therefore
\begin{equation} 
N_{2} = N_{1}-(2L-4). 
\label{eq:N2} 
\end{equation} 
Both \(N_{1}\) and \(N_{2}\) are obtained deterministically from the lattice size \(L\), 
without performing a heuristic search. 

For comparison, 
we apply \texttt{find\_embedding} to the same intermediate Hamiltonians. 
We try to embed the intermediate Hamiltonian for $L$ 
in $Z_m$ for $m = \lfloor (L+1)/2 \rfloor$, 
while the proposed method successfully supports the embedding 
of this particular combination of Hamiltonian and graph sizes.
Note that 
we can embed the parity Hamiltonian on a square lattice for $L = 24$, 
which has 576 parity spins, 
in a currently available quantum annealer with $Z_{12}$, 
provided that the embedding avoids inactive qubits in the actual device.
Since \texttt{find\_embedding} is a heuristic algorithm, 
different executions do not necessarily produce the same embedding, 
even when the problem and hardware graphs are unchanged. 
We therefore execute \texttt{find\_embedding} 400 times for each value of \(L\). 
From these trials, 
we first verify whether a valid embedding is obtained.
We then calculate the average number of used physical qubits, 
denoted by $\Nfave$, 
and the minimum number, 
denoted by \(N_{\mathrm{f,min}}\), 
over all successful trials.
In practice, 
embedding was successful in all trials; 
therefore, 
the average and the minimum were calculated over all trials.

\begin{figure}[t]
\begin{center}
\includegraphics[width=1\columnwidth]{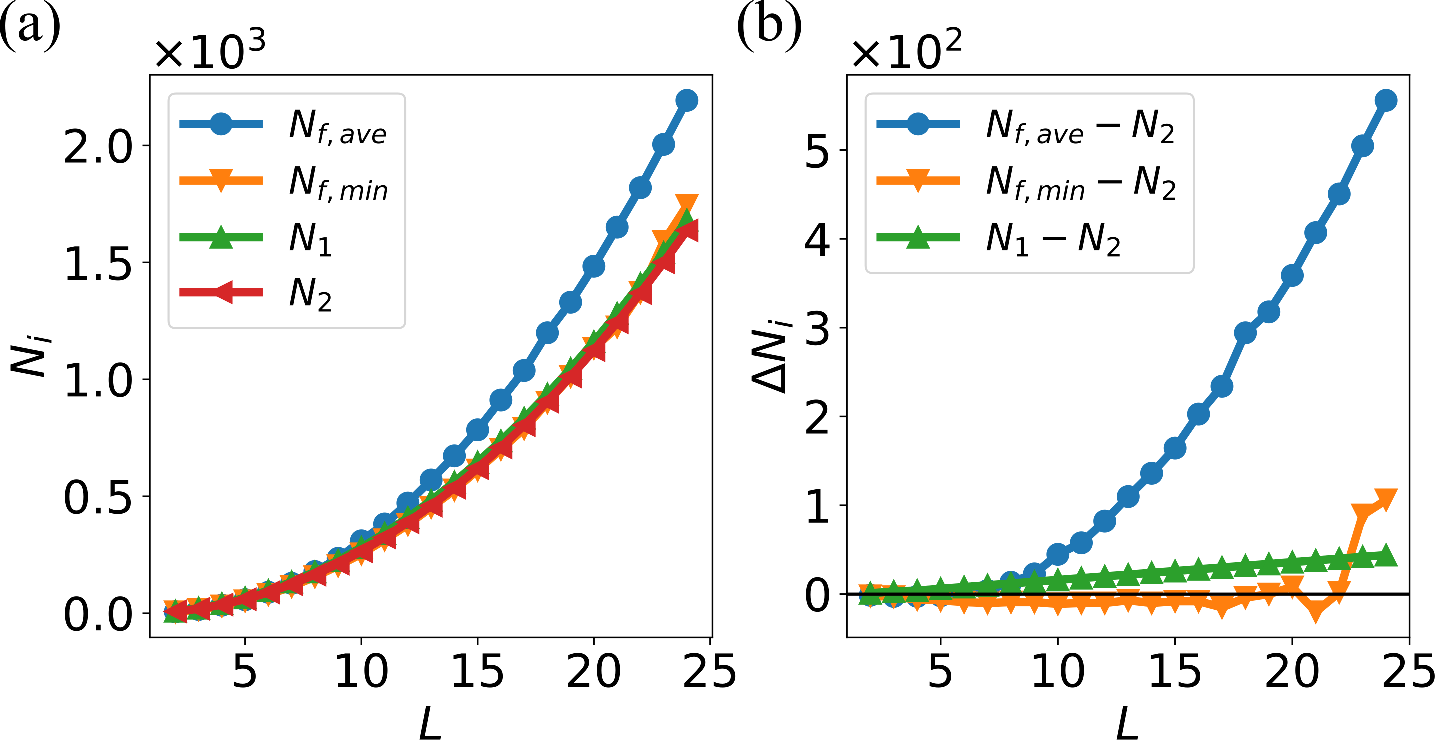}%
\caption{(a)The number $N_i$ of physical qubits 
		used by the embeddings as a function of side length $L$ of square lattices
		for the parity Hamiltonians.
		The intermediate Hamiltonians transformed 
		from the parity Hamiltonian for $L$
		are embedded in the Zephyr graph $Z_{m}$ 
		for $m = \lfloor (L+1)/2 \rfloor$.
		$N_1$ is for the proposed embedding 
		developed in Sec.~\ref{sec:embedding}.
		$N_2$ is for its variant discussed in Appendix.
		$\Nfave$ and $\Nfmin$ are the average and minimum numbers 
		of used physical qubits, respectively, 
		over 400 trials of \texttt{find\_embedding} for each value of $L$.
		(b) The differences of used physical qubits from $N_2$, 
		defined by $\Delta N_i = N_i - N_2$ 
		for $N_i = N_1$, $\Nfave$, and $\Nfmin$, 
		as functions of $L$.
		}
\label{fig:comparison}
\end{center}
\end{figure}

Figure~\ref{fig:comparison}(a) compares the number \(N_i\) of physical qubits 
used by the embeddings as a function of \(L\).
All four curves $N_1, N_2, \Nfave$, and $\Nfmin$ exhibit a similar overall dependence on the lattice size. 
However, 
the numbers \(N_{1}\) and \(N_{2}\) obtained using the proposed schemes 
are consistently smaller than 
the average value $\Nfave$ obtained using \texttt{find\_embedding}. 
On the other hand, 
\(N_{1}\) and \(N_{2}\) are close to the minimum value $\Nfmin$ 
observed in our repeated \texttt{find\_embedding} trials. 
To examine these differences more clearly, 
we define 
\begin{equation} 
\Delta N_i = N_i-N_{2}, 
\label{eq:delta_N} 
\end{equation}
which measures the additional number of physical qubits 
relative to the boundary-reduced variant of the proposed embedding. 
Figure~\ref{fig:comparison}(b) shows \(\Delta N_i\) as a function of \(L\). 
The relatively large value of \(\Delta \Nfave\) demonstrates that 
\texttt{find\_embedding} typically requires more physical qubits than the proposed schemes. 
Thus, 
although the heuristic algorithm can occasionally produce a qubit-efficient embedding, 
its typical output has a larger embedding overhead. 
Figure~\ref{fig:comparison}(b) also shows that the numbers of physical qubits 
used by the proposed schemes 
are close to the observed minimum $\Nfmin$, 
particularly for small \(L\). 
For \(L>20\), however, 
\(\Delta \Nfmin\) increases with \(L\) 
and eventually exceeds \(\Delta N_{1}\). 
This behavior indicates that 
repeated executions of \texttt{find\_embedding} can identify an embedding 
requiring nearly the same number of physical qubits 
as the proposed schemes 
when the problem size is sufficiently small. 
Nevertheless, as \(L\) increases, 
a larger number of repetitions is required 
to obtain such a high-quality embedding. 
It therefore becomes increasingly difficult for the heuristic search 
to find a qubit-efficient embedding for large problem instances. 

The obtained result shows that 
the proposed deterministic schemes typically use fewer physical qubits 
than a single typical execution of \texttt{find\_embedding}, 
while achieving a qubit count close to, or smaller than, 
the minimum found through repeated heuristic trials. 
However, 
note that 
we have confirmed that 
\texttt{find\_embedding} is capable of 
embedding the Hamiltonians of the same size 
into the graphs of the same size 
as those considered in the proposed method.

\subsection{Computational cost}

The computational cost provides another important distinction 
between the two approaches. 
In the present problem setting, 
the numbers of edges in both the problem graph and the hardware graph are proportional to their respective numbers of vertices. 
Under this condition, 
where both graphs are sparse, 
the computational time required by \texttt{find\_embedding} 
to obtain one possible embedding scales as 
$O(N_p^{2})$~\cite{J.Cai2014arXiv}, 
where \(N_p\) denotes the number of parity spins in the intermediate Hamiltonian. 
Moreover, 
obtaining an embedding that uses fewer physical qubits 
generally requires multiple executions of the heuristic algorithm. 
As indicated by the results in Fig.~\ref{fig:comparison}(b), 
the number of repetitions needed to obtain a high-quality embedding increases 
as the problem size grows. 
In contrast, 
both variants of the proposed scheme are deterministic. 
Once the lattice coordinates of the spins are specified, 
the corresponding physical qubits are assigned directly 
according to the mapping rules introduced above. 
An efficient embedding is therefore obtained in a single execution, 
without repeatedly searching for alternative embeddings. 
Because the assignment is performed once for each parity spin, 
the computational time of the proposed construction scales as
$O(N_p)$, 
which is lower than the scaling of the heuristic algorithm in the present setting. 
The advantages become particularly significant for large \(L\), 
for which qubit-efficient embeddings are essential 
and high-quality heuristic embeddings become increasingly difficult to obtain.



\section{Conclusion}

In this paper, 
we have presented a qubit-efficient embedding scheme 
for parity-encoded Hamiltonians in the SLHZ framework~\cite{N.Sourlas2005, W.Lechner2015} 
for quantum annealers whose qubit connectivity is given by the Zephyr graph $Z_m$~\cite{K.Boothby2021}. 
We have mapped a parity Hamiltonian [Eq.~(\ref{eq:H_parity})] containing multi-body interactions 
to an intermediate Hamiltonian $H'$ [Eq.~(\ref{eq:H_intermediate})] with only one- and two-body interactions. 
We have then provided an explicit constructive embedding [Eqs.~(\ref{eq:mapping}) and (\ref{eq:mapping_2})], 
where two-qubit chains in $Z_m$ are assigned to spins in $H'$ 
according to the systematic rule with translation operators 
that generate embeddings over the graph. 
The validity of the construction follows from the translational symmetry of the graphs and the explicit characterization of chain-to-chain connectivity 
that realizes all couplings required by $H'$.
Moreover, the embedding admits practical optionality.
We can choose between qubits within ancillary-spin chains and also swap chain roles to avoid inactive qubits in $Z_m$
while preserving the necessary couplings.
Changing the parity-Hamiltonian within logically equivalent representations provides an additional mechanism 
to avoid inactive qubits and also leads to an efficient use of a limited physical-qubit layout.
We have also demonstrated the high efficiency of the scheme.
The proposed embedding typically uses fewer physical qubits 
and has a lower computational-time scaling 
than a known heuristic algorithm, \texttt{find\_embedding}.
The advantages become particularly significant for large \(L\), 
for which qubit-efficient embeddings are essential.

Our embedding can serve as one of the key elements for benchmarking quantum annealing performance under the SLHZ scheme with available quantum annealers.
The scheme requires fewer qubits without introducing an imbalance in the number of physical qubits per parity spin.
This feature enables the implementation of relatively large-scale models and is unlikely to cause unnecessary bias in the behavior of the effective qubits.
As mentioned above, 
the optionality of our embedding 
would be useful when implementing quantum annealing on limited-scale devices including inactive qubits.

These benefits should be validated by implementing quantum annealing using the proposed embedding.
The previous study for the embedding in the Pegasus graph did not show a definite difference in the quantum annealing performance 
between two different embedding schemes~\cite{M.Cattelan2025}.
However, this comparison has been performed only for small-size problems.
Our method is expected to exhibit similar behavior for small problems, 
but it is not yet clear how the behavior would differ for larger problem sizes.
The proposed qubit-efficient embedding may contribute to improving the quality of solutions obtained through quantum annealing for large problem instances.
This point should be demonstrated using an actual quantum-annealing machine, 
but experimental validation is beyond the scope of this paper and is therefore left as future work.

Another important direction for future research is 
the generalization and abstraction of the present embedding approach 
to enable its application to other graph structures.
Such extensions would be valuable 
for establishing a more universal embedding framework. 


\begin{acknowledgment}
This paper is based on results obtained from 
Project No. JPNP16007 commissioned by the New Energy and Industrial Technology Development Organization (NEDO), Japan. 
The author acknowledges the use of Microsoft Copilot 
for assistance with language editing and improving the clarity of the manuscript. 
All scientific content and analysis were developed and verified by the author.
\end{acknowledgment}

\appendix
\section{A variant of embedding}
\label{sec:variant}
\begin{figure}[t]
\begin{center}
\includegraphics[width=1\columnwidth]{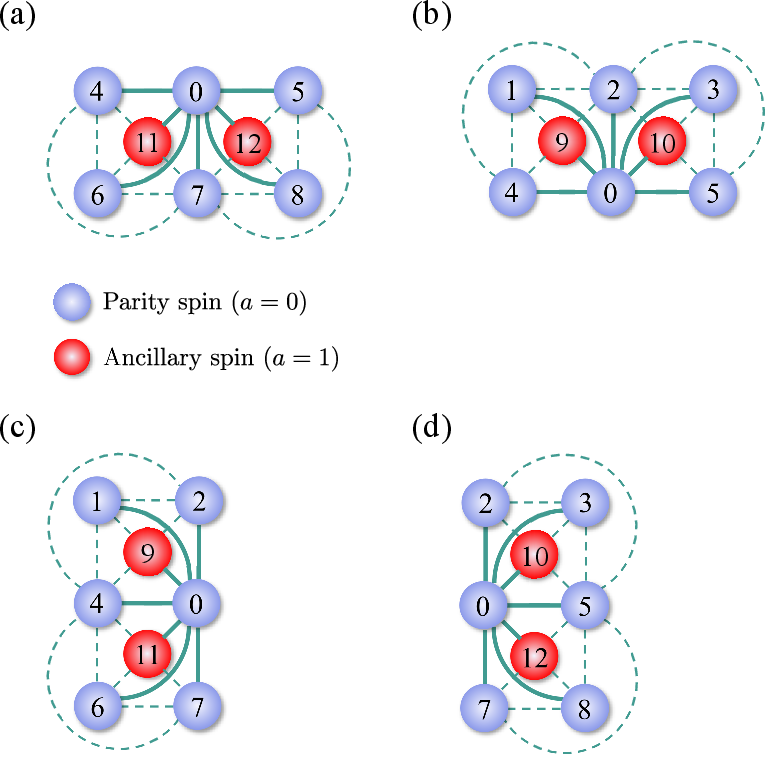}%
\caption{Spins of the intermediate Hamiltonian at the (a) top, (b) bottom, (c) right, and (d) left boundaries of the graph.
Spin 0 that we focus on and spins coupling to spin 0 are depicted, 
and the other spins are omitted for simplicity.
Edges for couplings to spin 0 are depicted as solid lines, 
while those for the other couplings are depicted as dashed lines.
These graphs are subgraphs of the graph shown in Fig.~\ref{fig:mapping}(a), 
and the labels of spins are consistent with those used in Fig.~\ref{fig:mapping}(a).
}
\label{fig:boundaries}
\end{center}
\end{figure}

At the boundaries of the intermediate-Hamiltonian graph, 
a parity spin has fewer couplings than a parity spin in the interior of the graph, 
as shown in Fig.~\ref{fig:boundaries}. 
Consequently, 
not all physical qubits in every boundary chain are necessarily required 
to realize the couplings of the intermediate Hamiltonian. 
Some two-qubit chains assigned to boundary parity spins can therefore be reduced 
to single physical qubits 
while preserving all the required couplings between chains. 
In this Appendix, 
we consider only the reduction of parity-spin chains. 
The reduction of ancillary-spin chains has already been discussed in Sec.~\ref{sec:optionality}. 

We impose the condition that, 
when one qubit is removed from a parity-spin chain, 
no qubit is simultaneously removed from a chain coupled to that parity-spin chain. 
In particular, 
two mutually coupled chains cannot both be reduced, 
because such a simultaneous reduction can eliminate a physical coupler 
required to realize the interactions with the surrounding chains. 
Under this condition, 
whether a parity-spin chain can be reduced is determined 
by examining whether one of its two physical qubits retains all the couplings 
required by the intermediate Hamiltonian. 

\begin{table}
\caption{Reduction of chains to single qubits for parity spins of the intermediate Hamiltonian 
		at the boundaries of the graph. 
		The subgraphs at the boundaries are depicted in Fig.~\ref{fig:boundaries}, 
		where the parity spin whose chain is reduced to a single qubit is labeled 0, 
		and the spins coupled to spin 0 are also shown.
		The second column gives the operator acting on the chain for the parity spin 
		to map its associated ancillary spin to a chain.
		For mapping $A$, the assigned chains for spins coupled to spin 0 are summarized in Table~\ref{table:correspondence}, 
		while Table~\ref{table:correspondence_2} is for mapping $A'$.
		These tables also summarize the correspondence to Table~\ref{table:chain_couplings}, 
		where parameters $u$ of coupled qubits are listed. 
		Taken together, 
		labels of chains for the coupled spins given in Table~\ref{table:chain_couplings} are obtained 
		as in the third column.
		Inspection of all rows corresponding to the labels confirms that 
		the required couplings of chains can be implemented 
		solely using the qubit characterized by parameter $u$ given in the fourth column.
		The chain for spin 0 is assumed to be parameterized as $(x_Z, y_Z, \alpha, \beta)$. 
		$\bar{\alpha}=1-\alpha$.
		}
\label{table:reduction}
\centering
\begin{tabular}{l c c c}
\hline
Boundary [Fig.~\ref{fig:boundaries}] & Mapping & Table \ref{table:chain_couplings} & Remaining-qubit's $u$\\
\hline
(a) Top & $A$ & 1, 4, 9, 6, 10, 11 & $\alpha$ \\
(b) Bottom & $A'$ & 1, 2, 3, 5, 8, 9, 10 & $\bar{\alpha}$ \\
(c) Right & $A'$ & 1, 2, 8, 6, 7, 10 & $\alpha$ \\
(d) Left & $A$ & 1, 2, 3, 9, 10, 11 & $\bar{\alpha}$ \\
\hline
\end{tabular}
\end{table}

Table~\ref{table:reduction} summarizes the physical qubit 
that can remain in a reduced parity-spin chain at each of the four boundaries. 
The entries are obtained by comparing the couplings 
required by the intermediate Hamiltonian 
with the chain-to-chain connectivity summarized in Table~I. 
We first explain the reduction at the top boundary in detail 
and then extend the same argument to the other boundaries. 
Consider spin 0 at the top boundary shown in Fig.~\ref{fig:boundaries}(a). 
We assume that 
the chain assigned to the corresponding ancillary spin, 
labeled 12 in the figure, 
is obtained by applying the operator \(A\) defined in Eq.~(\ref{eq:A}) 
to the chain assigned to spin 0. 
The spins coupled to spin 0 are labeled \(4,5,6,7,8,11,\) and \(12\) 
in Fig.~\ref{fig:boundaries}(a). 
By comparing Figs.~\ref{fig:chain_couplings}, \ref{fig:mapping}(b), and \ref{fig:boundaries}(a), 
the chains assigned to these spins can be identified with chains \(1,4,9,6,10,\) and \(11\) 
in Table~\ref{table:chain_couplings} and Fig.~\ref{fig:chain_couplings}. 
Although seven coupled spins are present, 
only six distinct chain labels appear in this correspondence 
because the chains assigned to spins 7 and 11 can both correspond to chain 10 in Table~\ref{table:chain_couplings}. 
The chain labels in Table~\ref{table:chain_couplings} are applicable to both \(\alpha=0\) and \(\alpha=1\). 
We next determine which physical qubit should remain in chain 0. 
The two physical qubits in a chain are distinguished by the Zephyr coordinate \(u=0\) or \(1\). 
The fourth column of Table~\ref{table:chain_couplings} specifies the values of \(u\) 
for the qubits in the neighboring chains 
that are coupled to chain 0. 
In this construction, 
chain-to-chain couplings are realized using internal couplers of the Zephyr graph, 
which connect physical qubits with opposite orientations. 
Thus, for example, 
a qubit with \(u=0\) in one of the neighboring chains shown in Fig.~\ref{fig:chain_couplings} couples to the qubit with \(u=1\) in chain 0. 
For chains \(1,9,10,\) and \(11\), 
both constituent qubits couple to chain 0, 
as indicated in the fourth column of Table~\ref{table:chain_couplings}. 
Therefore, the couplings to these chains can be retained by either physical qubit in chain 0. 
For chains 4 and 6, by contrast, 
only the qubit with $u=\bar{\alpha}$ in the neighboring chain couples to chain 0. 
Because the internal couplers connect qubits with opposite values of \(u\), 
these couplings are incident on the qubit with \(u=\alpha\) in chain 0. 
Recall that $\bar{\alpha} = 1 - \alpha$.
It follows that 
the \(u=\alpha\) qubit in chain 0 is coupled to all the distinct neighboring chains required at the top boundary. 
Chain 0 can therefore be reduced to the single qubit with \(u=\alpha\), 
while the qubit with \(u=\bar{\alpha}\) can be removed. 

The assumption concerning the associated ancillary spin is important for this reduction. 
When the ancillary-spin chain is obtained by applying \(A\) to the parity-spin chain, 
the single qubit with \(u=\alpha\) realizes all the required couplings at the top boundary. 
If, instead, a parity spin whose corresponding ancillary-spin chain is obtained by applying \(A'\) is selected at the same boundary, 
neither constituent qubit of the parity-spin chain individually possesses all the required couplings. 
In that case, 
the chain cannot be reduced to a single qubit. 
Therefore, 
whether a boundary parity-spin chain can be reduced depends both on the boundary at which the spin is located 
and on whether its associated ancillary-spin chain is generated by \(A\) or \(A'\). 

The same analysis can be applied to the bottom, right, and left boundaries shown in Figs.~\ref{fig:boundaries}(b), (c), and (d), respectively. 
At each boundary, 
one can identify a constituent qubit 
that retains all the physical couplers required by the intermediate Hamiltonian. 
The results are summarized in Table~\ref{table:reduction}. 

\begin{figure}[t]
\begin{center}
\includegraphics[width=1\columnwidth]{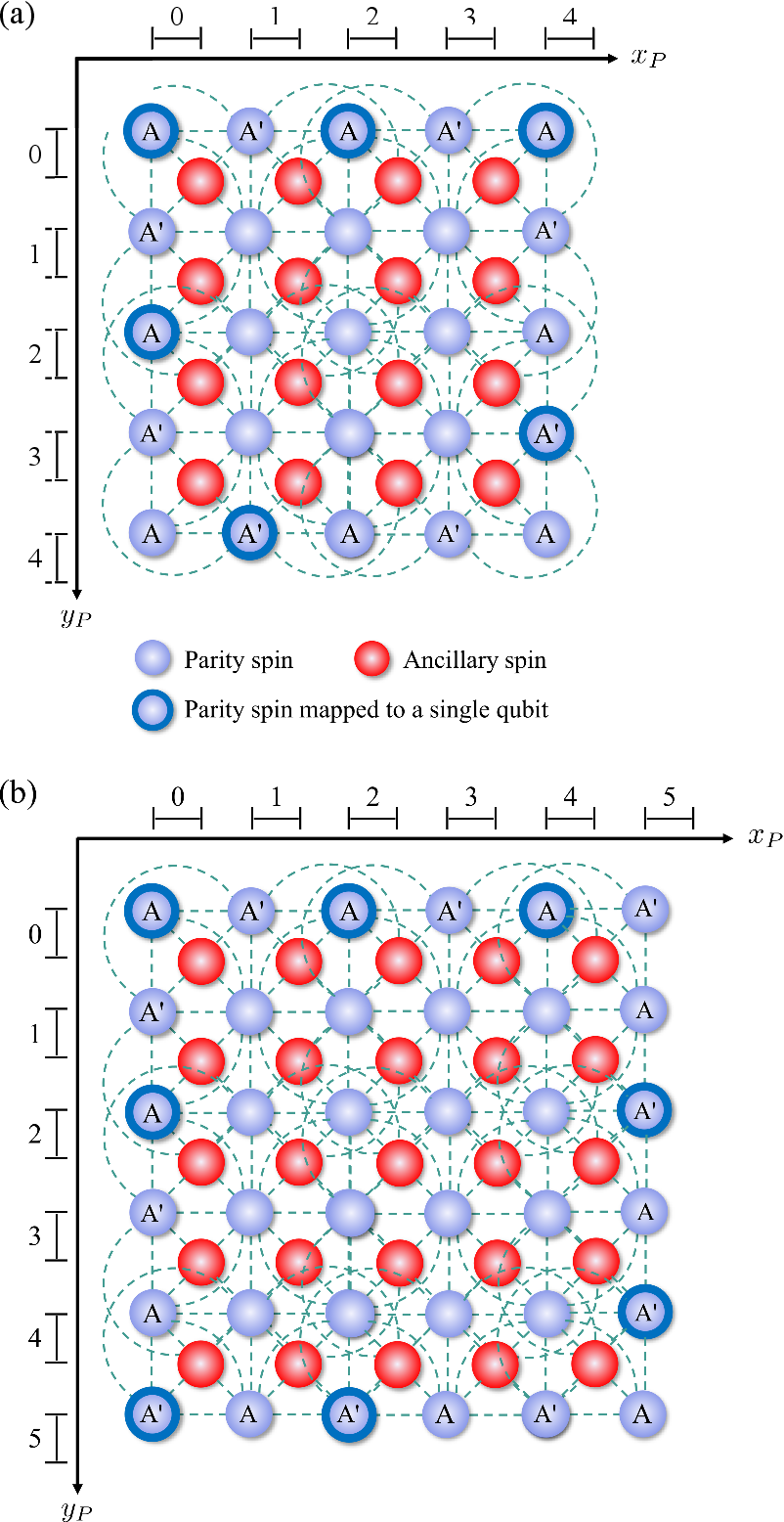}%
\caption{Examples of selection of parity spins mapped to a single qubit 
		for (a) odd and (b) even numbers of parity spins on each side of the graph.
		Parity spins enclosed with thick circles are selected to be mapped to a single qubit.
		No two adjacent spins are not selected simultaneously.
		$A$ and $A'$ on the parity spins at the boundaries indicate the operators 
		acting on their chains to map their associated ancillary spins to chains.
		The relationship between these operators and the selection is summarized 
		in Table~\ref{table:reduction}
		}
\label{fig:selection}
\end{center}
\end{figure}

We next determine a set of boundary chains that can be reduced simultaneously. 
Figure~\ref{fig:selection}(a) gives an example for a square lattice with an odd number of parity spins along each side. 
In accordance with the boundary conditions derived above, 
we select parity spins whose corresponding ancillary-spin chains are generated by \(A\) at the top and left boundaries. 
At the bottom and right boundaries, 
we instead select parity spins whose corresponding ancillary-spin chains are generated by \(A'\). 
The selections must also satisfy the condition that 
two coupled parity-spin chains are not reduced simultaneously. 
Accordingly, 
two parity spins that couple to each other cannot both be included in the set of reduced chains. 
In the example shown in Fig.~\ref{fig:selection}(a), 
the parity spins at \((0,4,0)\), \((3,4,0)\), and \((4,1,0)\) are not selected 
because their selection would violate this condition. 
The restriction affects the selection only in the vicinity of three corners of the square lattice. 
Away from these corners, 
the boundary chains satisfying the appropriate \(A\) or \(A'\) condition can be reduced 
according to the alternating pattern shown in the figure. 

We finally count the number of parity-spin chains that can be reduced in this manner. 
Let \(L\) denote the number of parity spins along each side of the square lattice. 
If the restriction arising from mutually coupled chains is temporarily ignored, 
a total of \(L\) parity-spin chains can be selected from the top and bottom boundaries combined. 
This result follows because the parity spins associated with \(A\) and \(A'\) alternate along each boundary, 
and the same alternating pattern occurs at both the top and bottom boundaries. 
For the same reason, 
a total of \(L\) parity-spin chains can be selected from the left and right boundaries combined. 
The preliminary count from the four boundaries is therefore \(2L\). 
However, 
the parity spin at the upper-left corner is included in both the top-boundary selection and the left-boundary selection 
and must be counted only once. 
This double counting reduces the number by one. 
In addition, 
the three parity spins identified above cannot be selected 
because of the condition prohibiting the simultaneous reduction of mutually coupled chains. 
The total number of parity-spin chains that can be reduced is consequently $2L-4$.
Each selected two-qubit chain is replaced by one physical qubit, 
so this process decreases the total number of physical qubits by \(2L-4\), 
as used in Eq.~(\ref{eq:N2}). 
The same number of parity-spin chains can be reduced when the lattice has an even number of parity spins along each side. 
An explicit selection for the even-\(L\) case is shown in Fig.~\ref{fig:selection}(b). 
Although the detailed alternating pattern differs from that for odd \(L\), 
the boundary conditions and the restriction on mutually coupled chains lead to the same result, 
namely \(2L-4\) reducible parity-spin chains.


\end{document}